\documentclass[11pt]{iopart}
\pdfoutput=1
\usepackage{graphics}
\usepackage{longtable}
\usepackage{graphicx}
\graphicspath{{figures/}}
\usepackage{amsfonts}
\usepackage{epsfig}
\usepackage{color}
\usepackage{amssymb}
\usepackage{subfigure}
\usepackage{caption}
\usepackage{geometry}
\usepackage{layout}
\usepackage{multicol}
\usepackage{arydshln}
\usepackage{float}
\usepackage[normalem]{ulem}

\expandafter\let\csname equation*\endcsname\relax
\expandafter\let\csname endequation*\endcsname\relax
\usepackage{amsmath}

\usepackage[dvipsnames]{xcolor}

\usepackage{tikz}
\DeclareRobustCommand\full  {\tikz[baseline=-0.6ex]\draw[thick] (0,0)--(0.5,0);}
\DeclareRobustCommand\dotted{\tikz[baseline=-0.6ex]\draw[thick,dotted] (0,0)--(0.54,0);}
\DeclareRobustCommand\dashed{\tikz[baseline=-0.6ex]\draw[thick,dashed] (0,0)--(0.54,0);}
\DeclareRobustCommand\chain {\tikz[baseline=-0.6ex]\draw[thick,dash dot] (0,0)--(0.5,0);}
\newcommand{\tikzcircle}[2][red,fill=red]{\tikz[baseline=-0.5ex]\draw[#1,radius=#2] (0,0) circle ;}

\usepackage{multirow}
\usepackage{parskip}      % space line between paragraphs
\usepackage{indentfirst}  % indentation at first paragraph

\usepackage{afterpage}    % blank page
\usepackage[font=scriptsize]{caption} % figure script small

\usepackage{pdfpages}   % insert a pdf page

\usepackage{graphicx,subfigure}

\usepackage{longtable}    % table divided into two pages

\usepackage{threeparttable}

\usepackage{threeparttablex} % for "ThreePartTable" environment
\usepackage{booktabs}        % for well-spaced horizontal rules

\usepackage{hyperref}
\hypersetup{
colorlinks=true,
linkcolor=blue,
citecolor=blue,    
urlcolor=black,
}

\usepackage{cleveref}

\usepackage{cite}

\begin{document}

\title[]{Atomic oxygen densities in He/O$_2$ micro-scaled atmospheric pressure plasma jets: a systematic model validation study}
\author{Youfan He$^1$, Ralf Peter Brinkmann$^1$, Efe Kemaneci$^1$, Andrew R. Gibson$^{2,3}$}

\address{$^1$ Institute of Theoretical Electrical Engineering, Faculty of Electrical Engineering and Information Technology, Ruhr University Bochum, Germany}
\address{$^2$ Research Group for Biomedical Plasma Technology, Faculty of Electrical Engineering and Information Technology, Ruhr University Bochum, Germany}
\address{$^3$ York Plasma Institute, School of Physics, Engineering and Technology, University of York, United Kingdom}

\ead{heyoufan1993@gmail.com, efekemaneci@gmail.com, andrew.gibson@york.ac.uk}

\begin{abstract}
\noindent 
Reactive species produced by atmospheric pressure plasma jets have high application potential in the fields of biomedicine and surface processing. 
An extensive validation between the simulation results in this work and measurement data from various research groups is carried out in order to reliably understand the complicated chemical kinetics defining the reactive species densities.
Atomic oxygen densities in parallel plate radio frequency driven He/O$_2$ micro-scaled atmospheric pressure plasma jets ($\mu$APPJs) have been measured in the literature by several research groups with different methods including: two-photon absorption laser induced fluorescence (TALIF) spectroscopy and optical emission spectroscopy (OES)-based methods. These measurement data with a variation of the absorbed power, the He gas flow rate and the O$_2$ mixture ratio are simulated in this paper with a zero-dimensional (0-D) plasma-chemical plug-flow model coupled with a two-term Boltzmann equation solver. The simulated atomic oxygen densities agree well with most of the measured ones. Specifically, good agreement is achieved between the simulations and most of the TALIF measurements over a range of operating conditions.  
Our model prediction accuracy relative to these TALIF measurements is quantified by the percentage error between the measured and simulated atomic oxygen densities. An approximate normal distribution is observed in the histogram plot of the percentage error, and the mean is close to zero. The mean is shifted positively and negatively in the case of removing a dominant atomic oxygen gain and loss reaction channel, which implies the underestimation and overestimation of the simulation results relative to the measurement data, respectively. 
This indicates that proper incorporation of the dominant reaction channels in the simulations plays a key role in the model prediction accuracy.
\\


\end{abstract}

\maketitle

\section{Introduction}
\label{sec:int}

A large variety of studies have been conducted on atmospheric pressure plasma jets \cite{Win15,Pen15,Fan17,Reu18,Lu21,Vie22} for their application prospects in the fields such as surface processing (e.g. etching, deposition) and biomedicine (e.g. bacteria inactivation, wound healing) \cite{Pen15,Fan17,Reu18}, as well as gas conversion (e.g. CO$_2$ dissociation, NH$_3$, NO synthesis) \cite{Urb18,Ste20,Du21,Du21_2,Ver24,Yu24}.
Correspondingly, diverse configurations of atmospheric pressure plasma jets were developed \cite{Win15,Pen15,Fan17,Lu21}, and wide range of fundamental experimental and computational studies were carried out \cite{Reu18,Lu21,Vie22}. 
Atmospheric pressure plasma jets are well suited to efficiently convert molecular gases into diverse reactive species, which are considered to play a major role in related applications such as bacteria inactivation \cite{Win15,Pen15} and polymer modification \cite{Fan17,Reu18}.
It was reported that the produced reactive atomic oxygen species is of key importance for the treatment of cancer cells \cite{Kim10_2} and polymer etching \cite{Fri12,Wes16,Wes16_2}. 
Furthermore, it was predicted in our previous work \cite{He21} that atomic oxygen has a significant influence on the production and destruction of nitric oxide, which has been suggested to be important in wound healing applications \cite{Gra12}.
Therefore, a detailed study of the atomic oxygen density produced by atmospheric pressure plasma jets is important to improve the performance of the corresponding applications.

Reactive species densities are affected by the complex plasma chemistry, which consists of diverse interactions between neutrals and ions. A fundamental understanding of these rich chemical kinetics is necessary for the development and optimization of reactive species production. 
Numerous detailed reaction sets have been established, such as the He/H$_2$O sets by Liu $et$ $al$ \cite{Liu10_FromEfe} and Schröter $et$ $al$ \cite{Sch18_PCCP}, the He/O$_2$ sets by Liu $et$ $al$ \cite{Liu10} and Turner \cite{Tur15}, the He/Air sets by Murakami $et$ $al$ \cite{Mur12,Mur13,Mur14} and Sun $et$ $al$ \cite{Sun19}, and the Ar/air set by Van Gaens $et$ $al$ \cite{Gae13,Gae14,Gae14_2}. The reliability of the established plasma-chemical models are generally assessed by the validation between the simulated and corresponding measured reactive species densities. For instance, the validations were conducted for the densities of atomic oxygen and hydroxyl radicals \cite{Sch18_PCCP}, helium metastable and reactive oxygen species \cite{Mur12}, ozone \cite{Gae13}, atomic oxygen and nitric oxide \cite{Gae14_2}, atomic oxygen \cite{Vas24} as well as argon metastable, ozone and nitrogen oxide species \cite{Sch15}.

The simulation results of the modellers during validation are mainly compared with the measurement data from a single type of plasma source from the collaborating experimental colleagues, and sometimes compared with those from diverse types of plasma sources in the literature. In the case of comparison with the data from a single type of plasma source, the inconsistency between the input parameters of the simulations and measurements is minimized by the close collaboration between the modellers and their experimental colleagues. In the case of comparison with the data from diverse types of plasma sources, the influence of the model limitations and experimental errors on the deviation between the simulation results and measurement data can be reduced. 
The comparison with the data from a single type of and diverse types of plasma sources provides the advantages of both approaches. 
In this case of validation between the simulations and measurements, the aforementioned inconsistency, influence of the model limitations and experimental errors are mitigated. Consequently, the reliability of the established chemical model is further improved by this case of validation. However, such validation is still lacking in the literature to the knowledge of the authors.

The development and optimization of reactive species production is one of the main research goals of parallel plate micro-scaled atmospheric pressure plasma jets ($\mu$APPJs), which have the advantages of the miniature and simple design for an easy and safe operation and a localized treatment \cite{Gol16}. 
The atomic oxygen densities produced by $\mu$APPJs has been of interest for a number of years, and multiple publications from various groups reported corresponding density measurements \cite{Was10,Bib11,Wes16,Rie20, Mye21, Ste21, Ste22, Win22}. 
These radio frequency driven $\mu$APPJs \cite{Was10,Bib11,Wes16,Rie20, Mye21, Ste21, Ste22, Win22} work on a similar principle, e.g. the feed gas flow is perpendicular to the electric field generated between two parallel planar electrodes. Many of these $\mu$APPJs are related to the COST-Jet, i.e. the European COST (Cooperation in Science and Technology) Reference Microplasma Jet \cite{Gol16}. Specifically, these $\mu$APPJs are the COST-Jet \cite{Rie20, Mye21, Ste21, Ste22}, the COST-Jet prototypes \cite{Was10,Bib11,Wes16}, and a capillary jet device (the mechanical and electrical design and operating principle are similar to those of the COST-Jet) \cite{Win22}. More details of these $\mu$APPJs are provided in section \ref{sec:setup}. It is worth to explicitly note that accurate measurements of the absorbed power in the plasma (one of the most important input parameter for the simulations) can be provided by the recent experimental studies of the $\mu$APPJs \cite{Wes16,Rie20, Mye21, Ste21, Ste22, Win22}. Therefore, the inconsistency between input parameters of the simulations and measurements can be reduced compared to earlier studies where the measurement of absorbed power was not regularly performed. Furthermore, the COST-Jet \cite{Rie20, Mye21, Ste21, Ste22} was developed as a reference source in accordance with a series of prototype sources used in the years before. The non-reproducibility of the COST-Jet was minimized by Golda $et$ $al$ \cite{Gol16} with a large amount of effort such as the refinement of the mechanical and electrical design. A reliable validation between the experiments and simulations is enabled by these reproducible measurement data of atomic oxygen densities.

The prediction accuracy of the simulated reactive species densities relative to the corresponding measurement data is one of the key issues for the plasma modelling. Quantification of the prediction accuracy was reported by several publications \cite{Tur15,Tur16,Ber17,Wan18}, i.e. the influence of the uncertainty of the simulation input parameters (especially a large number of reaction rate coefficients that are subject to error) on the simulation output results. It was indicated by Turner \cite{Tur15,Tur16} that the uncertainty of the rate coefficients in the He/O$_2$ chemical model results in the uncertainty of the calculated species densities. Such a density uncertainty is a factor of two to five in most cases, however it is even more than a factor of ten in some extreme cases \cite{Tur15}. It was presented by Berthelot $et$ $al$ \cite{Ber17} and Wang $et$ $al$ \cite{Wan18} that different combinations of rate coefficients based on their uncertainties are used to predict the uncertainty of the simulation output results such as uncertainty of electron density and temperature (15$\%$) in the CO$_2$ plasma \cite{Ber17}, uncertainty of CO$_2$ and CH$_4$ conversion (24$\%$ and 33$\%$, respectively) \cite{Wan18}. Furthermore, the complex chemical model (almost 400 reactions) for the conditions relevant to biomedical applications was simplified (about 50 reactions), and in the sensitivity analysis around 10 reaction rate coefficients were found to cause most of the uncertainty \cite{Tur16}. The reaction rate coefficients primarily responsible for the uncertainty in the relevant model prediction are also identified in \cite{Ber17,Wan18}. Attention to these critical rate coefficients, i.e. resulting in the uncertainty, can contribute to a huge improvement in the model prediction accuracy. 

Good prediction accuracy can be achieved by including the key reactions in the chemical model as comprehensively as possible.
Recently, the Quantemol database developed by Tennyson $et$ $al$ \cite{Ten22} has been devoted to providing data on all the relevant chemical kinetics that could be important in the plasmas. For instance, a fast algorithm was developed by Hanicinec $et$ $al$ \cite{Han20} to output the key reaction set defining the species density of interest, and a regression model was trained by the same group \cite{Han23} on available reaction and species data extracted from the multiple databases to fast approximate the unknown rate coefficients of the involved chemical kinetics. The former focuses on filtering out the key reactions, while the latter provides the opportunity to find the potentially important reactions.

The prediction accuracy can be confirmed by the agreement between the simulation results and measurement data, as noted earlier \cite{Sch18_PCCP,Mur12,Gae13,Gae14_2,Vas24,Sch15}.
However, the presented agreement was typically validated by a qualitative assessment. Specifically, the simulation results and the measurement data under certain operating conditions were simultaneously shown in a figure, and an agreement was assessed qualitatively instead of with quantitative number.
In other words, discussion of the prediction accuracy in terms of quantification of the agreement between the simulated and measured species densities is still lacking in the literature of plasma modelling. 
Such a quantification can be potentially obtained by the quantitative metrics.
Recently, in other areas of computational chemistry, multiple metric approaches for purpose of quantifying the prediction accuracy were reported by Vishwakarma $et$ $al$ \cite{Vis21}. These metrics can be used to quantitatively assess the quality, reliability and applicability of a given model, and to further compare the performance of the different models. For example, the mean deviation between the simulation results and a lager number of measurement data under diverse operating conditions from various groups can be quantitatively assessed, and the influence of an absence of the key chemical reactions on the species density of interest can be quantitatively evaluated.

It should be emphasized that the prediction accuracy as a function of the deviation between the simulations and measurements is determined by diverse factors such as model limitation, experimental error, and potential inconsistency between the input parameters of the simulations and measurements.
The influence of the model limitation on the prediction accuracy can be mitigated to some extent by attention to uncertainty of the critical rate coefficients \cite{Tur15,Tur16,Ber17,Wan18} and by focus on filter of the key reactions \cite{Ten22,Han20,Han23}, as discussed above about the work of the shown publications. In the current work, the main aim is to quantitatively assess the prediction accuracy of the model used, with respect to atomic oxygen densities, through systematic comparison with a large range of experimental data sets using similar plasma sources and diverse measurement methods. 

The focuses of this study are summarized as the following two points:
\begin{itemize}
\item In section \ref{sec:validation}, the atomic oxygen density simulation results of this work are validated against the measurement data from multiple publications, which were conducted under different operating conditions, e.g. the absorbed power, the gas flow rate and the mixture ratio.
\item In section \ref{sec:statistik}, the prediction accuracy of the model relative to the atomic oxygen densities measured by TALIF is quantified by the percentage error between the simulation results and measurement data. 
The influence of removing the dominant atomic oxygen gain and loss reaction channels from the model on the prediction accuracy of the model is quantitatively revealed.  
\end{itemize}
The $\mu$APPJs simulated in this paper are discussed in section \ref{sec:setup}. The pseudo one dimensional plug-flow model providing the one dimensional spatially resolved simulation results along the gas flow direction used in this study is described in section \ref{sec:mod}. The considered plasma-chemical reaction scheme is described in section \ref{sec:Chem_kin}. The key results are summarized in section \ref{sec:con}.

\section{Experimental model system}
\label{sec:setup}

The $\mu$APPJs \cite{Was10,Bib11,Wes16,Rie20, Mye21, Ste21, Ste22, Win22} are simulated in this study, i.e. the COST-Jet \cite{Rie20, Mye21, Ste21, Ste22}, the COST-Jet prototypes \cite{Was10,Bib11,Wes16}, and a subsequent evolution design of the COST-Jet \cite{Win22}. 
Specifically, the atomic oxygen densities in the plasma channel and in the near effluent (close to the exit of the plasma channel) of the aforementioned radio frequency driven He/O$_2$ $\mu$APPJs \cite{Was10,Bib11,Wes16,Rie20, Mye21, Ste21, Ste22, Win22} measured with different methods in the multiple publications of various groups are simulated with the theoretical approach and the chemical kinetics provided in sections \ref{sec:mod} and \ref{sec:Chem_kin}, respectively. The rectangular plasma channels of these $\mu$APPJs are designed as a cross-field configuration, i.e. the generated electric field between the two parallel planar electrodes is perpendicular to the feed gas flow. The feed gas is injected through one side of the two smallest cross sections, and exhausted from the other side. Good optical access for species density measurements is provided by the two side glass plates. For further details regarding the structures of these $\mu$APPJs, see \cite{Was10,Bib11,Wes16,Rie20, Mye21, Ste21, Ste22, Win22}.

In section \ref{sec:res}, the $\mu$APPJ of West \cite{Wes16}, as a prototype, was investigated to inform the construction and operation of the COST-Jet \cite{Gol16}. The $\mu$APPJs in the recent studies reported by Riedel $et$ $al$ \cite{Rie20}, Myers $et$ $al$ \cite{Mye21} and Steuer $et$ $al$ \cite{Ste21,Ste22} are the COST-Jet. The $\mu$APPJ provided by Winzer $et$ $al$ \cite{Win22} is similar to the COST-Jet. However, they are slightly different, e.g. dielectric glasses were additionally assembled on the electrode surfaces confining the plasma in \cite{Win22} to prevent a glow-to-arc transition at high absorbed power values so that the plasma is still stable at higher powers. 
The absorbed power value in the plasma is available in the experimental report of the aforementioned $\mu$APPJs \cite{Wes16,Rie20, Mye21, Ste21, Ste22, Win22}. The absorbed power, which is important for the electron-impact processes in \ref{sec:chemkin} such as ionization and excitation mechanisms, can be directly used as an input parameter of the simulations to minimize the potential modelling error (e.g. due to a transfer efficiency of 5$\%$ between the generator input power and plasma absorbed power assumed for the $\mu$APPJs \cite{Was10,Bib11} simulated in our previous study \cite{He21}). It should be emphasized that the gas impurity and power uncertainty were reported as major reasons for the irreproducible experimental results of $\mu$APPJs \cite{Gol16}. A large amount of effort was conducted by Golda $et$ $al$ \cite{Gol16} for the COST-Jet to avoid the gas impurity (e.g. with the sealing improvement) and avoid the uncertainty of measured absorbed power (e.g. with integrated probes). Therefore, the potential errors as a result of the inconsistency between the input parameters of simulations and measurements are further minimized. This prompts a more precise prediction of the simulation results to the measurement data. The comparisons between our simulation results and the measurement data of the $\mu$APPJs \cite{Wes16,Rie20, Mye21, Ste21, Ste22, Win22} are shown and discussed in section \ref{sec:res}. 
Note that the atomic oxygen densities produced by similar $\mu$APPJs were also reported by other literature, e.g. the density measurement data in the effluent by Willems $et$ $al$ \cite{Wil19} and those in the plasma channel under a peak-to-peak voltage value of 500 V by Korolov $et$ $al$ \cite{Kor21_2}. However, such $\mu$APPJs (e.g. \cite{Wil19, Kor21_2}) are not simulated in this work attributing to our focus on the density in the plasma channel and the necessity of knowing the absorbed power used in the simulations for this work.

In \ref{sec:He21APPJ}, the $\mu$APPJs reported by Waskoenig $et$ $al$ \cite{Was10} and Bibinov $et$ $al$ \cite{Bib11} are the old versions of the COST-Jet \cite{Gol16}. A power transfer efficiency of 5$\%$ for converting the generator input power (provided by these two old versions) to the absorbed power in the plasma was assumed in the simulation results of our previous study \cite{He21} in order to accurately predict the measurement data in \cite{Was10,Bib11}. These two $\mu$APPJs are simulated in this study with the same operating conditions reported in \cite{He21}, but simulated with the plug-flow model in section \ref{sec:mod} and simulated with the updated and supplemented chemical kinetics in this work given in section \ref{sec:Chem_kin} and \ref{sec:chemkin}. Note that the corresponding simulation results shown in \ref{sec:He21APPJ} are only used as a comparison with those in our previous study \cite{He21} due to the assumption of the 5$\%$ power transfer efficiency. These results are not analyzed in section \ref{sec:res}.

Operating conditions of the above-mentioned He/O$_2$ $\mu$APPJs \cite{Was10,Bib11,Wes16,Rie20, Mye21, Ste21, Ste22, Win22} from various publications of different groups are summarized in table \ref{tab:OperationCondition} including the plasma channel size, the He gas flow rate, the O$_2$ mixture ratio, the absorbed power in the plasma, the position of the atomic oxygen density measurements and the corresponding method used for the density measurements. 
Precise plasma channel pressure values were not provided in \cite{Wes16,Rie20,Mye21, Ste21, Ste22, Win22}, therefore $1\times10^5$ Pa is used in our simulations for the $\mu$APPJs shown in section \ref{sec:res}. The gas temperatures used in the simulations of these $\mu$APPJs are estimated from the experimentally defined relationships between the effluent gas temperature and absorbed power provided in \cite{Rie20,Win22}. Note that \cite{Rie20} (P. 5) provided the relationship at z = 3 mm (i.e. in the effluent and 3 mm distance from the plasma channel exit) in an absorbed power range of 0.2 - 1.0 W, while \cite{Win22} (P. 6-7) provided the relationship at z = 0 mm (i.e. at the plasma channel exit) in an absorbed power range of 0.5 - 6.0 W. The relationship of gas temperature versus absorbed power is roughly linear shown both by \cite{Rie20} and \cite{Win22}. The gas temperatures at z = 3 mm \cite{Rie20} are overall smaller than those at z = 0 mm \cite{Win22} around 5 - 30 K in the absorbed range of 0.06 - 6.50 W. Averaged values of the aforementioned gas temperature data between \cite{Rie20} and \cite{Win22} at the considered absorbed powers are estimated in this study to be the following fit function:
\begin{equation}
{\rm T}_{{\rm g}}{\rm(K)}= 302.6591 + 34.4318 \: {\rm P}_{{\rm abs}}{\rm(W)}, \: (0.06 {\rm W} \leqslant {\rm P}_{{\rm abs}} \leqslant 6.50 {\rm W}),
\label{eqn:Tg_Pabs}
\end{equation}
where ${\rm T}_{{\rm g}}$ in K is the gas temperature, and ${\rm P}_{{\rm abs}}$ in W is the absorbed power in the plasma. These estimated temperature values are assumed to be the volume-averaged gas temperatures in the plasma channel region for z $<$ 0 mm, which are smaller than the gas temperatures at the plasma channel exit for z $=$ 0 mm in \cite{Win22}. The above-mentioned fit function is used to calculate the gas temperatures for an absorbed power range of 0.06 - 6.50 W, which are used in the simulations of the $\mu$APPJs \cite{Wes16,Rie20,Mye21, Ste21, Ste22, Win22} shown in table \ref{tab:OperationCondition} and section \ref{sec:res}. In addition, it is observed in our simulations (not shown here) that the atomic oxygen densities as a function of power are invariant between the simulations of the $\mu$APPJ \cite{Wes16} for a varying gas temperature with equation (\ref{eqn:Tg_Pabs}) and those for a constant of 350 K. In other word, the atomic oxygen densities are not sensitive to the gas temperature under the considered operating conditions. Furthermore, an approximately constant gas temperature for a varying O$_2$ mixture ratio was experimentally reported by the He/O$_2$ $\mu$APPJ \cite{Win22} (P. 6).  
Following the values in our previous study \cite{He21}, a gas temperature of 345 K and a plasma channel pressure of $1\times10^5$ Pa are adopted in our simulations for the $\mu$APPJ used in reference \cite{Was10}, while 370 K and $101 \: 325$ Pa are adopted for the $\mu$APPJ used in reference \cite{Bib11} shown in \ref{sec:He21APPJ}.

The simulated atomic oxygen densities in this work are compared with the corresponding measured ones using different methods: including two-photon absorption laser induced fluorescence (TALIF) spectroscopy \cite{Was10,Wes16,Rie20, Mye21, Ste21, Ste22}, helium state enhanced actinometry (SEA) \cite{Ste22, Win22} and optical emission spectroscopy (OES) \cite{Bib11}.
In order to better understand the potential deviation between the simulation results and measurement data, it is of importance to be aware of the fundamental principles of the different measurement methods and the corresponding investigation region. Furthermore, due to the focus on simulations in this work, only a brief introduction to the three measurement methods mentioned above is given below. 

In TALIF, the energy of two laser photons is used to excite a ground state atom of interest \cite{Wes16} (P. 29). The effective decay rate (i.e. the reciprocal of the life time) of the excited atomic species is mainly affected by two factors: radiative decay, and collisional quenching with other species. The fluorescence photon as a part of the radiative decay is emitted during the de-excitation of the atomic species, and subsequently a fluorescence signal is measured \cite{Wes16} (P. 34). Collisional quenching plays a more important role in the excited atomic species decay at higher pressures, and it is regarded as one of the largest possible sources of errors \cite{Wes16} (P. 32) during the evaluation of the aforementioned effective decay rate. 
The aforementioned effective decay rate is one of the most important parameters affecting the evaluation of the ground state atom density of interest \cite{Wes16} (P. 35,37). The absolute density of the ground state atom is calibrated by using the noble gas, e.g. the xenon gas used to calibrate the atomic oxygen density in the He/O$_2$ plasmas \cite{Was10,Wes16,Rie20, Mye21, Ste21, Ste22}. 

Picosecond (ps) laser systems have been used to directly measure the effective decay rate of the oxygen excited state in ps-TALIF approach \cite{Wes16,Rie20,Mye21}. Such decay rate can not be resolved by nanosecond (ns) laser pulses, and the effective lifetime is generally calculated on the basis of the gas mixture and known rate constants for collisional quenching in ns-TALIF approach \cite{Ste21,Ste22,Was10}. 
The accuracy of the effective lifetime calculation for the oxygen excited state in the effluent region is not guaranteed due to the uncertain quenching rate coefficients and potentially unknown species concentrations \cite{Wes16} (P. 37), especially during mixing with ambient air \cite{Wes16,Rie20,Mye21}. In comparison, the calculation accuracy of the effective decay rate in the plasma channel region is in a reasonable degree as a result of the controlled feed gas\cite{Wes16} (P. 37), e.g. the dominant background gas He and defined O$_2$ admixtures \cite{Ste21,Ste22,Was10}. Therefore, ps-TALIF has advantages for measurements in the effluent region \cite{Wes16,Rie20,Mye21}, while ns-TALIF should be better-suited for those in the plasma channel region \cite{Ste21,Ste22,Was10}. 

The atomic oxygen density measured with the SEA approach \cite{Ste22,Win22} is based on the methods of the classical actinometry and energy resolved actinometry \cite{Gre14,Tsu17}. 
The density measured with the classical actinometry is determined from the intensity ratio of two spectral lines. One spectral line is from the gas to be studied (specifically from an excited oxygen state), and the other is from the actinometer gas of known density (typically from an excited argon state). Only direct electron-impact excitation is assumed in the classical actinometry approach to produce the oxygen and argon excited states from the respective ground states. The potentially important dissociative electron-impact excitation is neglected in the classical actinometry, but is considered in the energy resolved actinometry to improve the accuracy of the atomic oxygen density measurements \cite{Tsu17} (P. 2). A third spectral line (from another excited oxygen state) is introduced in the energy resolved actinometry, and this allows for the simultaneous measurement of both the atomic oxygen density and the local mean electron energy \cite{Gre14} (P. 2). 
In the SEA approach, the aforementioned third spectral line in the energy resolved actinometry is replaced with the spectral line from an excited helium state. Several optimizations are achieved in the SEA measurements compared to the energy resolved actinometry, e.g. the reduced experimental complexity and the improved precision of the measured mean electron energy.
In the OES approach used in \cite{Bib11}, the atomic oxygen density is determined from the spectral transition intensity, the cross-sections of the excitation processes, and the electron density as a function of the measured nitrogen molecular emission intensities.
For further details of the TALIF, SEA and OES approaches, see \cite{Was10,Bib11,Wes16,Rie20, Mye21, Ste21, Ste22, Win22}.

\begin{ThreePartTable}
{\scriptsize
\begin{longtable}[h]{p{3.3cm}p{2.1cm}p{1.5cm}p{1.5cm}p{1.9cm}p{2.1cm}p{2.2cm}}
\caption{
Operating conditions of the $\mu$APPJs simulated in this work. These $\mu$APPJs are from several publications of various groups \cite{Was10,Bib11,Wes16,Rie20, Mye21, Ste21, Ste22, Win22}. Atmospheric pressure of $1\times10^5$ Pa and gas temperatures as a function of absorbed powers (estimated from the effluent gas temperature measurements in \cite{Rie20,Win22}, see equation (\ref{eqn:Tg_Pabs}) in section \ref{sec:setup}) are used in our simulations for the $\mu$APPJs \cite{Wes16,Rie20, Mye21, Ste21, Ste22, Win22} shown in section \ref{sec:res}. Following the values in our previous study \cite{He21}, gas temperatures of 345 K and 370 K and plasma channel pressures of $1\times10^5$ Pa and $101 \: 325$ Pa are adopted in our simulations for the $\mu$APPJs in \cite{Was10} and in \cite{Bib11}, respectively, shown in \ref{sec:He21APPJ}.
}. 
\\\hline
\\[\dimexpr-\normalbaselineskip+3pt]
\\[\dimexpr-\normalbaselineskip+3pt]
References of $\;\;\;\;\;\;\;\;$ He/O$_2$ $\mu$APPJs  & Plasma channel size (mm$^3$)     &    He gas flow rate (sccm)    &    O$_2$ mixture ratio ($\%$)    &    Absorbed power (W)$\:^a$    &    measurement position (mm)$\:^b$    &    measurement method         \\ 
\\[\dimexpr-\normalbaselineskip+3pt]
\hline
\\[\dimexpr-\normalbaselineskip+3pt]
\endfirsthead

\\[\dimexpr-\normalbaselineskip+3pt]
\\[\dimexpr-\normalbaselineskip+7pt]
\textbf{{\underline{in section \ref{sec:res}}}}   
&        &        &        &        &        &        \\ 
\\[\dimexpr-\normalbaselineskip+5pt] 

West 2016 \cite{Wes16}    
&    $1 \times 1 \times 30$    &   1000   &     0.5   &     1.00 - 5.00      &   z $>$ 0 $\;\;\;\;\;\;\;\;$ (near effluent)  &  ps-TALIF    \\ 
\\[\dimexpr-\normalbaselineskip+3pt] 

Riedel $et$ $al$ 2020 \cite{Rie20}    
&    $1 \times 1 \times 30$    &   1000   &     0.5   &     0.06 - 1.25      &    z = 1  &  ps-$\:\&\:$ns-TALIF    \\ 
\\[\dimexpr-\normalbaselineskip+3pt] 

Myers $et$ $al$ 2021 \cite{Mye21}    
&    $1 \times 1 \times 30$    &   1000   &     0.1 - 1.0   &     0.75      &    z = 0  &  ps-TALIF    \\ 
\\[\dimexpr-\normalbaselineskip+3pt] 

Steuer $et$ $al$ 2021 \cite{Ste21}    
&    $1 \times 1 \times 30$    &   200 - 1200   &     0.5   &     1.00      &    z = -30 - 0  &  ns-TALIF    \\ 
\\[\dimexpr-\normalbaselineskip+3pt]

Steuer $et$ $al$ 2022 \cite{Ste22}    
&    $1 \times 1 \times 30$    &   1000   &     0.5   &     0.06 - 1.25$\:^a$     &   z = -15 or 0   &  ns-TALIF$\:\&\:$SEA    \\ 
\\[\dimexpr-\normalbaselineskip+3pt] 

Winzer $et$ $al$ 2022 \cite{Win22}    
&    $1 \times 1 \times 40$    &   1000   &     0.5   &     0.50 - 6.50     &   z = -20 or 0   &  SEA    \\ 
\\[\dimexpr-\normalbaselineskip+3pt] 

Winzer $et$ $al$ 2022 \cite{Win22}    
&    $1 \times 1 \times 40$    &   1000   &     0.1 - 2.0   &     1.00, 5.00     &   z = -20 or 0   &  SEA    \\ 
\\[\dimexpr-\normalbaselineskip+3pt] 

Winzer $et$ $al$ 2022 \cite{Win22}    
&    $1 \times 1 \times 40$    &   1000   &     0.5   &     1.00, 5.00     &    z = -40 - 0  &  SEA    \\ 
\\[\dimexpr-\normalbaselineskip+3pt]

\hline

\\[\dimexpr-\normalbaselineskip+3pt]
\\[\dimexpr-\normalbaselineskip+7pt]
\textbf{{\underline{in \ref{sec:He21APPJ}}}}   
&        &        &        &        &        &        \\ 
\\[\dimexpr-\normalbaselineskip+5pt]

Waskoenig $et$ $al$ 2010 \cite{Was10}    
&    $1 \times 1 \times 40$    &   995.0249   &     0.5   &     0.40 - 1.00     &   z = -20   &  ns-TALIF    \\ 
\\[\dimexpr-\normalbaselineskip+3pt] 

Bibinov $et$ $al$ 2011 \cite{Bib11}    
&    $1 \times 1.3 \times 40$    &   1500   &     1.5   &     1.5     &    z = -40 - 0  &  OES    \\ 
\\[\dimexpr-\normalbaselineskip+3pt] 

\hline

\label{tab:OperationCondition}
\end{longtable}
}

\begin{tablenotes}
\tiny
\item[]$^a$ Applied driving voltage instead of absorbed power was given in \cite{Ste22}. Since both $\mu$APPJs in \cite{Rie20} and \cite{Ste22} are the COST-Jet and were operated with the same conditions, the reported absorbed power of \cite{Ste22} in table \ref{tab:OperationCondition} is interpolated from the characteristics of absorbed power versus driving voltage measured in \cite{Rie20} (P. 5). 
\item[]$^b$ z $<$ 0 mm, z $=$ 0 mm and z $>$ 0 mm represent the region of plasma channel, plasma channel exit and plasma effluent, respectively. 

\end{tablenotes}

\end{ThreePartTable}

\section{Model}
\label{sec:mod}

Pseudo one dimensional plug-flow models have been reported in a number of publications \cite{He21,Mur14,Sch18_PCCP,Gae13,Gae14,Gae14_2,Sta04}. A pseudo one dimensional plug-flow model identical to that in our previous study \cite{He21} is used in this work. The only exception is that the wall loss of ions is not considered in this study due to their negligible influence on the simulation results \cite{He24} (P. 5). Therefore, only a brief summary regarding the model is provided as below. 
The model solves the species particle balance equations and an electron energy balance equation to obtain the time resolved plasma properties including species concentrations and effective electron temperature in an infinitesimal plug volume. It should be emphasized that this plug volume co-moves with the gas flow, therefore the time evolution of the plasma properties in this volume obtained from the balance equations is mapped to the one dimensional spatial position in the gas flow direction according to the gas flow velocity (see equation (5) in \cite{He21}). 
The aforementioned effective electron temperature is corresponding to the mean electron energy of a non-Maxwellian electron energy distribution function (EEDF) \cite{God93,LieBook2005}. The non-Maxwellian EEDF is self-consistently calculated by calling a Boltzmann solver, i.e. the open-source simulation tool Lisbon kinetics Boltzmann (LoKI-B) published by Tejero-del-Caz $et$ $al$ \cite{Tej19}. The solver is based on the steady-state solution of the Boltzmann equation under the two-term approximation. Note that the electron kinetics are mainly controlled by the background helium and oxygen densities, since they are dominant and virtually invariant under the considered operating conditions in this study. For the sake of reducing the simulation duration, the non-Maxwellian EEDF obtained from the Boltzmann solver for a corresponding steady-state plasma composition is used during the whole time evolution of the plasma properties in the plug-flow model. For further details of the aforementioned self-consistent calculation of the non-Maxwellian EEDF, see \cite{He21}.  
The plasma channel gas temperature used in the simulations is estimated from the experimentally defined relationship between the effluent gas temperature and absorbed power provided in \cite{Rie20,Win22}, see section \ref{sec:setup}.
For further details of implementing the plug-flow model, see \cite{He21}.

The plug-flow model in this work is limited to only properly calculate the plasma properties in the plasma channel region, while that in the effluent region is not implemented. One focus of this work is dedicated to establishing an accurate He/O$_2$ chemical kinetics (see section \ref{sec:Chem_kin}) through the reliable validation between the simulated atomic oxygen densities and measured ones from a number of publications by different groups \cite{Wes16,Rie20,Mye21,Ste21,Ste22,Win22} (see section \ref{sec:res}). 
However, the measurement data were collected from different positions: i.e. along the gas flow direction in the plasma channel region \cite{Ste21,Win22}, at the middle or exit of the plasma channel region \cite{Mye21, Ste22, Win22}, and in the near effluent region \cite{Wes16,Rie20}. 
It is computationally predicted in this work that the atomic oxygen density in the plasma channel region monotonically increases along the gas flow direction, and the plug-flow model calculation results at the middle of the plasma channel are close to those at the exit of the plasma channel for the considered operating conditions (see figures \ref{fig:Ste21} and \ref{fig:Win22}$(c)$). Furthermore, it was experimentally confirmed by Willems $et$ $al$ \cite{Wil19} (P. 4) and Myers $et$ $al$ \cite{Mye21} (P. 9) that the atomic oxygen density in the effluent region of the $\mu$APPJs monotonically decreases, and a proximity of the measurement data in the near effluent region to those at the plasma channel exit is observed. 
In other words, the atomic oxygen densities at the plasma channel exit are comparable with those at the middle of the plasma channel region and those in the near effluent region. 
Therefore, for the sake of consistency in section \ref{sec:res}, only the plug-flow model calculation results at the plasma channel exit are used to compared with the measurement data,  which were collected at the middle or exit of the plasma channel region \cite{Mye21, Ste22, Win22} and collected in the near effluent region \cite{Wes16,Rie20}. The plug-flow model calculation results along the gas flow direction in the plasma channel region are used to compare with the corresponding one dimensional spatially resolved measurement data \cite{Ste21,Win22}.

\section{Chemical kinetics}
\label{sec:Chem_kin}

\begin{table}[hbt!] \footnotesize
\centering
\caption{
The considered species in the $\mathrm{He/O_2}$ model.
}
\begin{tabular}{l}
\\[\dimexpr-\normalbaselineskip+3pt]
\hline
\\[\dimexpr-\normalbaselineskip+3pt]
\\[\dimexpr-\normalbaselineskip+3pt]
$ \mathrm{He/O_2}$ plasma:
$ \mathrm{He}$,
$ \mathrm{He(2 ^3S)}$,
$\mathrm{He_2^*}$,
$\mathrm{He^+}$,
$\mathrm{He_2^+}$,\\
\\[\dimexpr-\normalbaselineskip+3pt]
$\qquad\qquad\quad\:\:\:$ 
$\mathrm{O(^3 P)} $,
$\mathrm{O_2}$($v=0$),
$\mathrm{O_3}$,
$\mathrm{O(^1 D)}$,
$\mathrm{O_2 (a^1 \Delta_g)}$,
$\mathrm{O^+}$,
$\mathrm{O_2^+}$,
$\mathrm{O_4^+}$,
$\mathrm{O^-}$,
$\mathrm{O_2^-}$,
$\mathrm{O_3^-}$,
$\mathrm{O_4^-}$,\\
\\[\dimexpr-\normalbaselineskip+3pt]
$\qquad\qquad\quad\:\:\:$ 
$e$\\
$\qquad\qquad\quad\:\:\:$ 
$--------------------------------$\\
\\[\dimexpr-\normalbaselineskip+3pt]
$\qquad\qquad\quad\:\:\:$ 
Additionally considered species compared to our previous work \cite{He21}: $\mathrm{O_2 (b^1 \Sigma_{g^+})}$\\
\\[\dimexpr-\normalbaselineskip+3pt]
\hline
\end{tabular}
\label{tab:SpeciesHeO2}
\end{table}

The species considered in the $\mathrm{He/O_2}$ plasma are listed in table \ref{tab:SpeciesHeO2}. All the reaction channels included in this work are reported in \ref{sec:chemkin}. The $\mathrm{He/O_2}$ chemical model has been developed in our previous study \cite{He21}. Our simulation results of radio-frequency plasmas in \cite{He21} were benchmarked against the calculated electron density and electron temperature in \cite{Was10}. These simulation results were also validated against the measured atomic oxygen densities in \cite{Was10,Bib11}, the measured electron density and ozone density in \cite{Bib11}. 
In the current study, besides from the aforementioned two references, more measurement data of atomic oxygen density from a range of publications \cite{Wes16,Rie20, Mye21, Ste21, Ste22, Win22} are compared with our simulation results in order to further validate and optimize the developed $\mathrm{He/O_2}$ chemical model. For further details, see sections \ref{sec:setup} and \ref{sec:res}. 
It was predicted in \cite{He21} that the simulated atomic oxygen densities of the considered $\mu$APPJs are not affected by the vibrationally exited oxygen molecules included in the chemical model. Therefore, for the sake of reducing the simulation duration, these vibrationally exited states are not considered in the chemical kinetics of this study.

As a response to the need to optimize the $\mathrm{He/O_2}$ model, numerous updates and supplements are conducted in this study based on the chemical kinetics developed in our previous work \cite{He21}. All the updates and supplements are remarked in \ref{sec:chemkin}. For the sake of simplicity, only several key points are summarized as follows:
\begin{enumerate}
\item[(1)] 
The electron-impact reactions in tables \ref{tab:ReactionListHe}, \ref{tab:ReactionListHeO2}, \ref{tab:ReactionListHeO2updated}, \ref{tab:ReactionListHeO2updated_loki} and \ref{tab:elaHeO2} are incorporated in the 0-D model and LoKI-B solver according to the IST-Lisbon database, which yields good agreement between calculated and measured swarm parameters with the helium complete set \cite{IST20230227,Alv14} and also yields good agreement with oxygen complete set for E/N values between 10 and 1000 Td \cite{IST20231031,Alv14}. Only the cross-section data belonging to the complete set \cite{IST20230227,IST20231031,Alv14} are used in the solution to the Boltzmann equation \cite{Tej19} for cases marked with $ f(\epsilon) $ in the aforementioned tables, while those (not part of the complete set \cite{IST20230227,IST20231031,Alv14}) are directly evaluated to calculate rate coefficients according to the calculated EEDF for cases marked with $ f(\sigma) $ in these tables. The energy loss as a result of the excitation from ground state to vibrationally excited states and higher electronically excited states shown in table \ref{tab:ReactionListHeO2updated_loki} is considered since these excitation reactions are part of the complete set \cite{IST20230227,IST20231031,Alv14}. However, these vibrationally excited states and higher electronically excited states are not included in our $\mathrm{He/O_2}$ model due to the lack of corresponding chemical kinetics data. In other words, the production and destruction reaction channels of these excited states are not considered, and therefore the corresponding state densities are not calculated.
\item[(2)] 
The updates and supplements of the chemical kinetics in this work are based on a reaction mechanism for oxygen plasmas recently reported by Dias $et$ $al$ \cite{Dia23}, 
since it was shown that their 0-D simulation results are in good agreement with the measured $\mathrm{O(^3 P)} $, $\mathrm{O_2}$ electronically ground state, $\mathrm{O_2 (a^1 \Delta_g)}$ and $\mathrm{O_2 (b^1 \Sigma_{g^+})}$ densities in a DC glow discharge at low pressure. 
In order to better capture the production and destruction of $\mathrm{O(^3 P)} $, $\mathrm{O_2 (a^1 \Delta_g)}$ and $\mathrm{O_2 (b^1 \Sigma_{g^+})}$, an effort of updating and supplementing the corresponding reaction channels was conducted in \cite{Dia23}. In view of this, $\mathrm{O_2 (b^1 \Sigma_{g^+})}$ is additionally considered in our $\mathrm{He/O_2}$ model shown in table \ref{tab:SpeciesHeO2}. Furthermore, the rate coefficients in our previous $\mathrm{He/O_2}$ model \cite{He21} are updated to those in \cite{Dia23} for cases of the same reaction channels present in \cite{He21} and \cite{Dia23} shown in table \ref{tab:ReactionListHeO2}, 
and the reactions absent in \cite{He21} but present in \cite{Dia23} are supplemented into this study shown in table \ref{tab:ReactionListHeO2updated}. 
The neutral wall reactions in table \ref{tab:wrHeO2} are updated and supplemented in accordance with \cite{Dia23}, e.g. the wall recombination of the atomic oxygen $ \mathrm{O(^3P)} + \mathrm{wall} \rightarrow 1/2\mathrm{O}_2  $ is added into this study. This addition is based on a recent experimental study of Booth $et$ $al$ \cite{Boo20}, which was further discussed in \cite{Dia23}. 
It was reported in \cite{Dia23} (P. 12) and our previous study \cite{He21} (P. 10) that the probability of the atomic oxygen wall recombination used in the simulation results plays an important role in accurate prediction of the referenced measurement data. In other words, the atomic oxygen density is significantly affected by this wall recombination for the studied setup conditions reported in \cite{Dia23,He21}.
\item[(3)]
The partial supplements of the chemical kinetics shown in table \ref{tab:ReactionListHeO2updated} are conducted according to a simulation study of $\mathrm{He/O_2}$-containing plasma recently reported by Brisset $et$ $al$ \cite{Bri21}. Their $\mathrm{He/O_2}$ model was based on the work of Turner \cite{Tur15}, which made an effort to quantify the uncertainty of the predicted species densities due to the uncertainty of the reaction rate coefficients. We conduct the corresponding supplements in this work, e.g. the reactions regarding more interactions between helium and oxygen species.

\end{enumerate}

\section{Results}
\label{sec:res}

\subsection{Comparison between the simulated and measured atomic oxygen density}
\label{sec:validation}

\begin{figure}[h!]
\begin{minipage}[b]{0.5\textwidth}
\includegraphics[width=0.95\textwidth]{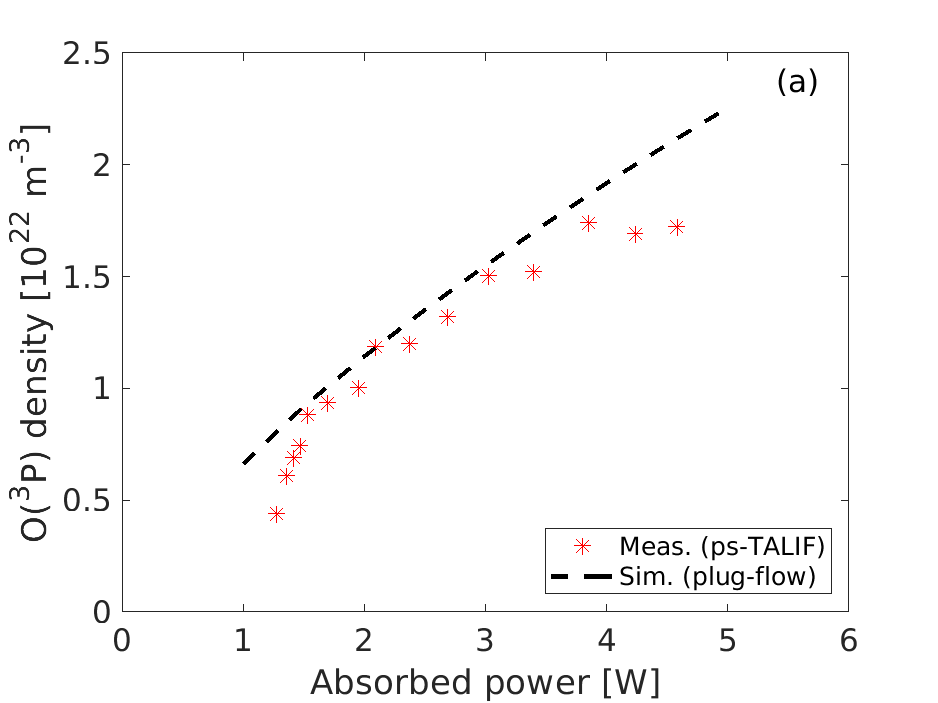} \\
\end{minipage}
\begin{minipage}[b]{0.5\textwidth}
\includegraphics[width=0.95\textwidth]{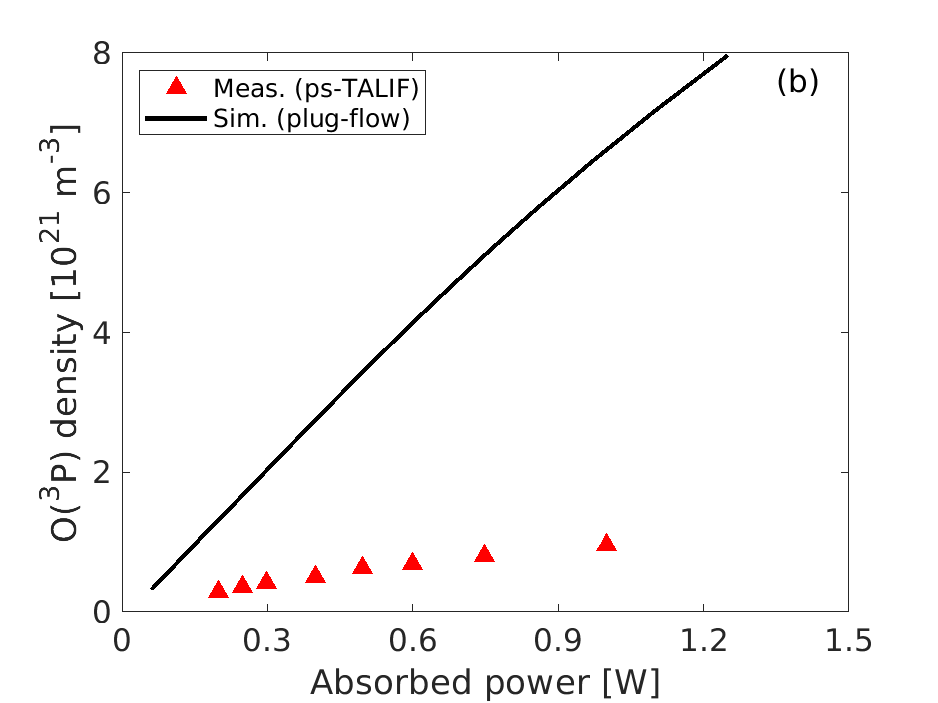} \\ 
\end{minipage}
\begin{minipage}[b]{0.5\textwidth}
\includegraphics[width=0.95\textwidth]{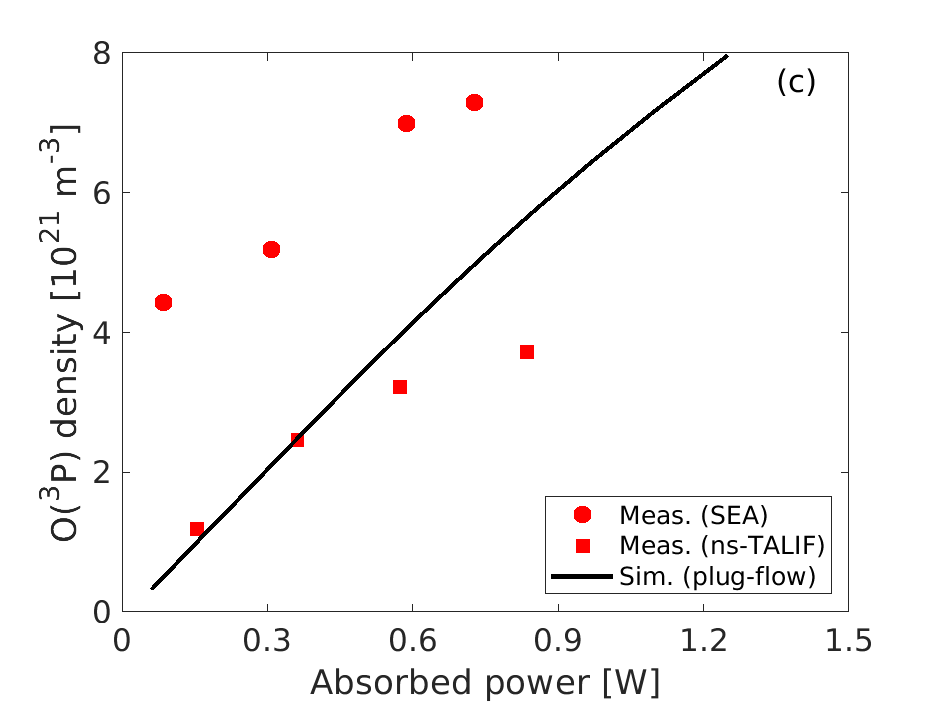} \\ 
\end{minipage}
\begin{minipage}[b]{0.5\textwidth}
\includegraphics[width=0.95\textwidth]{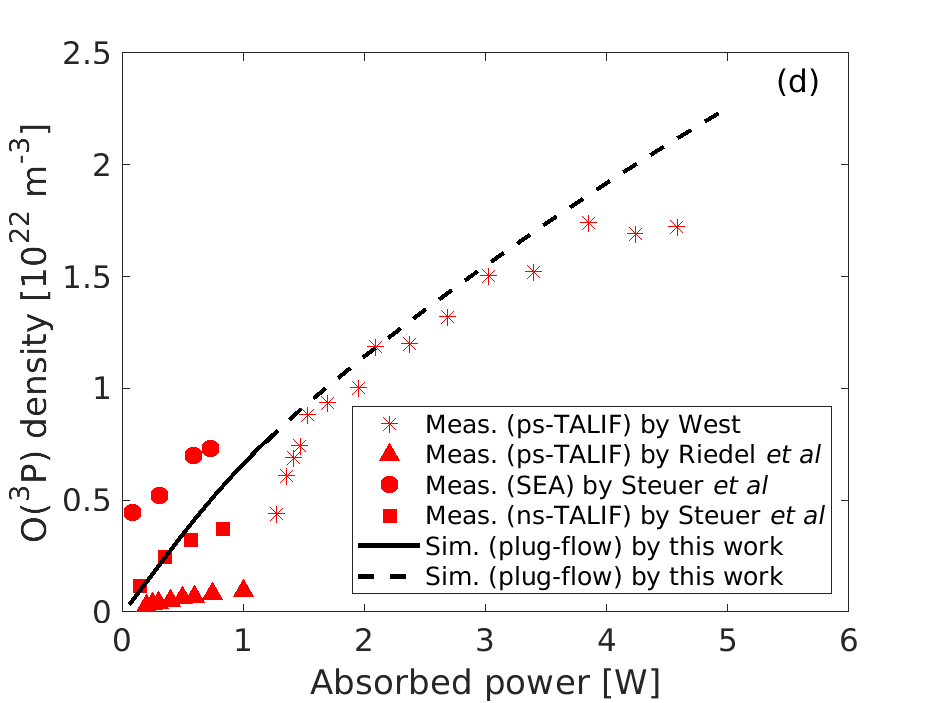} \\ 
\end{minipage}
\caption{
The $\mathrm{O(^3P)}$ densities of the He/$\mathrm{O_2}$ $\mu$APPJs for a variation of the absorbed power from 0.06 W to 5.00 W. (a) The ps-TALIF measurement data in the near effluent of the COST-Jet prototype reported by West 2016 \cite{Wes16} (\textcolor{red}{$\ast$}) and the plug-flow model calculations of this work (\textcolor{black}{\dashed}). (b) The ps-$\&$ns-TALIF measurement data in the effluent at 1 mm  distance from the plasma channel exit of the COST-Jet reported by Riedel $et$ $al$ 2020 \cite{Rie20} (\textcolor{red}{$\blacktriangle$}) and the plug-flow model calculations of this work (\textcolor{black}{\full}). (c) The ns-TALIF (\textcolor{red}{$\blacksquare$}) and SEA (\tikzcircle[red, fill=red]{2.0pt})  measurement data in the plasma channel of the COST-Jet reported by Steuer $et$ $al$ 2022 \cite{Ste22} and the plug-flow model calculations of this work (\textcolor{black}{\full}). 
The measurement data and model calculations provided in figures (a) - (c) are summarized in figure (d) for the sake of a straightforward comparison.
The shown plug-flow model calculations are the simulation results at the plasma channel exit, see section \ref{sec:mod}.
$1000$ sccm He gas flow mixed with $0.5\%$ O$_2$ are fed to the $1 \times 1 \times 30$ mm$^3$ plasma channel of the aforementioned $\mu$APPJs. 
}\label{fig:Wes16_Rie20_Ste22}
\end{figure}

The $\mathrm{O(^3P)}$ densities of the He/$\mathrm{O_2}$ $\mu$APPJs in a range of the absorbed power around from 0.06 W to 5.00 W measured by West 2016 \cite{Wes16}, Riedel $et$ $al$ 2020 \cite{Rie20} and Steuer $et$ $al$ 2022 \cite{Ste22} together with the corresponding plug-flow model calculations of this work are shown in figure \ref{fig:Wes16_Rie20_Ste22}. 
The plug-flow model calculation values are the simulation results at the plasma channel exit, see section \ref{sec:mod}.
The operating conditions of these $\mu$APPJs are provided in the figure caption, and more details such as the measurement methods and positions are described in section \ref{sec:setup}. 
It is found both by the measurements and simulations of the studied $\mu$APPJs that the $\mathrm{O(^3P)}$ densities increase monotonically with the increasing absorbed power. 
An agreement between the ps-TALIF measurement data of the COST-Jet prototype by West \cite{Wes16} and our simulation results in the absorbed power range of 1.00 - 5.00 W is obtained in figure \ref{fig:Wes16_Rie20_Ste22}(a). The measurement data are slightly smaller than the simulation results. This is attributed to that the measurements were conducted in the near effluent, while the simulations provide the values at the plasma channel exit.
It was also reported by similar $\mu$APPJs of Willems $et$ $al$ \cite{Wil19} and Myers $et$ $al$ \cite{Mye21} that the measured $\mathrm{O(^3P)}$ densities in the near effluent are slightly smaller than those at the plasma channel exit. 
However, our simulation results at the plasma channel exit overestimate the ps-$\&$ns-TALIF measurement data in the effluent at 1 mm distance from the plasma channel exit of the COST-Jet reported by Riedel $et$ $al$ \cite{Rie20}. The overestimation is around by a factor of 4.4 - 6.9 in the absorbed power range of 0.20 - 1.00 W presented in figure \ref{fig:Wes16_Rie20_Ste22}(b). 
Such a non-negligible overestimation may be partly due to the different simulation and measurement positions.
The ns-TALIF and SEA measurements were conducted by Steuer $et$ $al$ \cite{Ste22} in the plasma channel region of the COST-Jet. These measurement data are usually obtained at the middle or exit of the plasma channel, where both positions give the similar $\mathrm{O(^3P)}$ density under the considered operating conditions in this work (see figures \ref{fig:Ste21} and \ref{fig:Win22}(c)). The ns-TALIF measurement data are well captured by our simulation results at the plasma channel exit as shown in figure \ref{fig:Wes16_Rie20_Ste22}(c). 
Furthermore, our simulated $\mathrm{O(^3P)}$ density slightly underestimate the SEA measurement data, and our simulated mean electron energy (around 3 eV) is relatively close to the SEA measurement data (around 4.2 eV) in the absorbed power range of 0.08 - 0.72 W (not shown here). 
The similar underestimation is also predicted in the simulations of figure \ref{fig:Win22}. 
For a straightforward comparison, the measurement data and model calculations provided in figures \ref{fig:Wes16_Rie20_Ste22}(a) - (c) are summarized in figure \ref{fig:Wes16_Rie20_Ste22}(d). The overall trend and absolute values of the measurement data are captured by our model calculations.

\begin{figure}[h!]
\centering
\includegraphics[scale=0.50,clip=true,trim=0cm 0cm 0cm 0cm,]{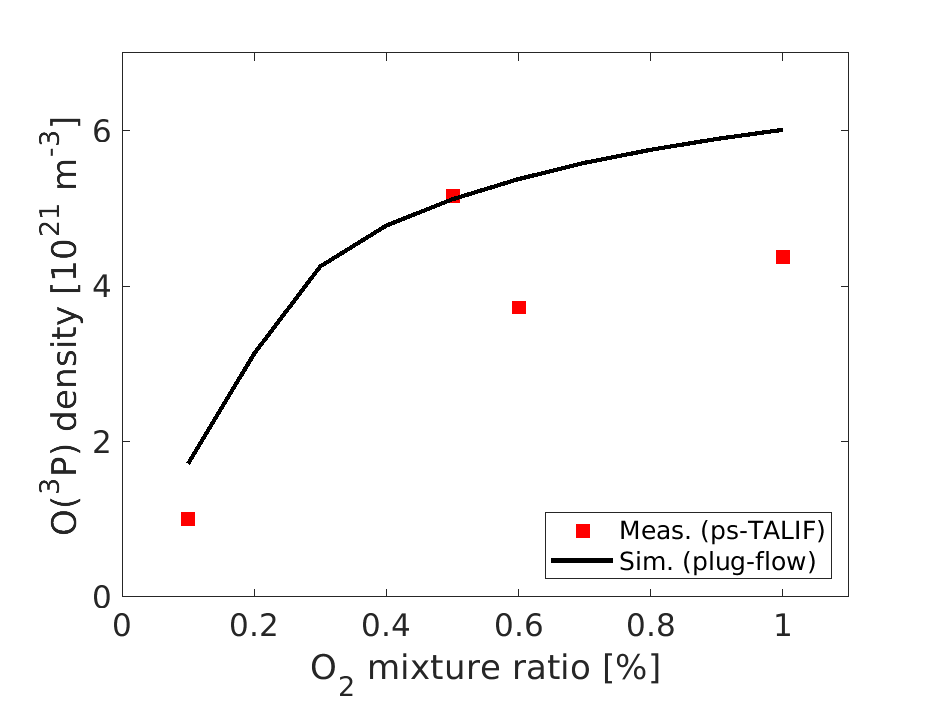}
\caption{
The $\mathrm{O(^3P)}$ density of the He/$\mathrm{O_2}$ $\mu$APPJ for a variation of the $\mathrm{O_2}$ mixture ratio from $0.1\%$ to $1.0\%$. The density at the plasma channel exit of the COST-Jet was measured with the ps-TALIF approach by Myers $et$ $al$ 2021 \cite{Mye21} (\textcolor{red}{$\blacksquare$}) and is simulated with the plug-flow model of this work (\textcolor{black}{\full}). $1000$ sccm He gas flow mixed with the depicted O$_2$ ratio are fed to the $1 \times 1 \times 30$ mm$^3$ plasma channel driven by 0.75 W absorbed power. 
}\label{fig:Mye21}
\end{figure}

The $\mathrm{O(^3P)}$ density at the plasma channel exit of the He/$\mathrm{O_2}$ $\mu$APPJ (i.e. the COST-Jet) in a range of the $\mathrm{O_2}$ mixture ratio from $0.1\%$ to $1.0\%$ measured by the ps-TALIF approach of Myers $et$ $al$ 2021 \cite{Mye21} together with the corresponding plug-flow model calculations of this work are shown in figure \ref{fig:Mye21}. The operating conditions are provided in the figure caption, and more details such as the measurement method and position are described in section \ref{sec:setup}. It is observed both by the measurements and simulations under the considered operating conditions that the $\mathrm{O(^3P)}$ density increases with increasing $\mathrm{O_2}$ mixture ratio, and starts to saturate at $1.0\%$ $\mathrm{O_2}$. Furthermore, similar measured and simulated $\mathrm{O(^3P)}$ densities are obtained.

\begin{figure}[h!]
\centering
\includegraphics[scale=0.50,clip=true,trim=0cm 0cm 0cm 0cm,]{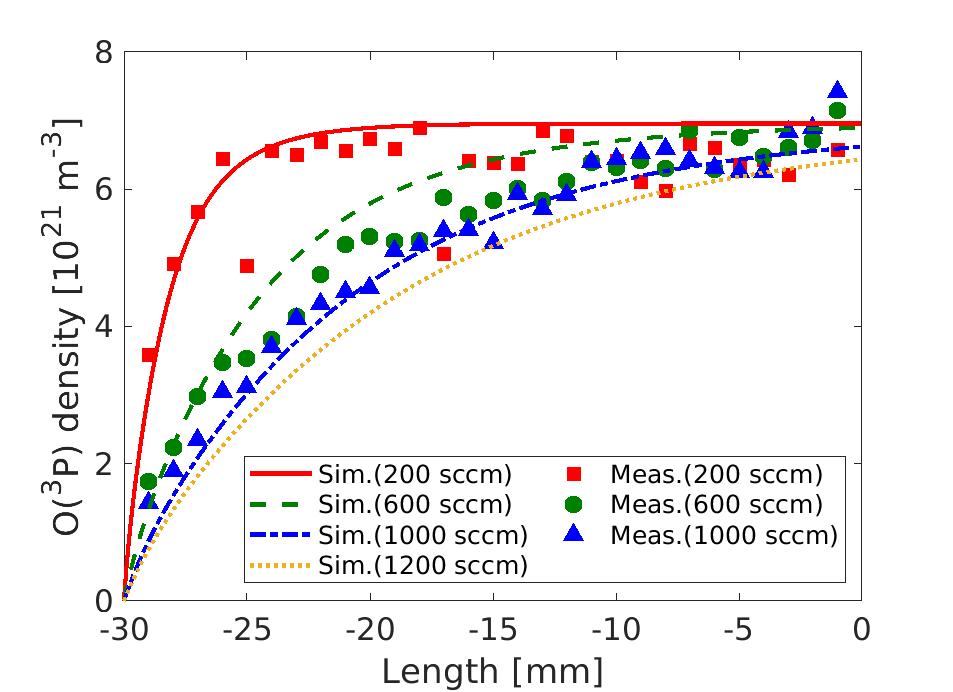}
\caption{
The one dimensional spatially resolved $\mathrm{O(^3P)}$ density along the gas flow direction in the plasma channel region of the He/$\mathrm{O_2}$ $\mu$APPJ for a variation of the feed 200 sccm, 600 sccm, 1000 sccm, and 1200 sccm He gas. The solid points (200 sccm \textcolor{red}{$\blacksquare$}, 600 sccm \tikzcircle[OliveGreen, fill=OliveGreen]{2.0pt}, and 1000 sccm \textcolor{blue}{$\blacktriangle$}) were measured with the ns-TALIF approach by Steuer $et$ $al$ 2021 \cite{Ste21}, and the density values of these solid points (i.e. the values at the middle between two electrodes) are obtained by averaging the measurement data at x=0.4 mm and x=0.6 mm of figure 2 in \cite{Ste21}, which were provided in \cite{RUBrdr_Ste21}. The lines (200 sccm \textcolor{red}{\full}, 600 sccm \textcolor{OliveGreen}{\dashed}, 1000 sccm \textcolor{blue}{\chain}, and 1200 sccm \textcolor{orange}{\dotted}) are simulated with the plug-flow model of this work. The depicted He gas flow mixed with $0.5\%$ O$_2$ are fed to the $1 \times 1 \times 30$ mm$^3$ COST-Jet plasma channel driven by 1.00 W absorbed power. 
}\label{fig:Ste21}
\end{figure}

The one dimensional spatially resolved $\mathrm{O(^3P)}$ density along the gas flow direction in the plasma channel region of the He/$\mathrm{O_2}$ $\mu$APPJ (i.e. the COST-Jet) for cases of feeding 200 sccm, 600 sccm and 1000 sccm He gas flow rate measured with the ns-TALIF approach by Steuer $et$ $al$ 2021 \cite{Ste21} together with the corresponding plug-flow model calculations of this work are shown in figure \ref{fig:Ste21}. The operating conditions are provided in the figure caption, and more details such as the measurement method and position are described in section \ref{sec:setup}. Note that the measurement data, representing the $\mathrm{O(^3P)}$ density along the gas flow direction at the middle between two electrodes, are obtained by averaging the measurement data at x=0.4 mm and x=0.6 mm of figure 2 in \cite{Ste21}, which were provided in \cite{RUBrdr_Ste21}. A good agreement between these measurement data and our model calculations is observed for a varying He gas flow rate. 
It is shown both by the measurements and simulations that for the lowest gas flow rate the $\mathrm{O(^3P)}$ density is saturated at earlier locations in the plasma channel, while for a higher gas flow rate the $\mathrm{O(^3P)}$ density is saturated only close to the plasma channel exit.
This is due to the smaller residence time of the species in the plasma channel at a higher gas flow rate. It is shown by the simulation results of 1200 sccm He gas flow rate that the $\mathrm{O(^3P)}$ density at the plasma channel exit is reduced by a further increasing gas flow.

\begin{figure}[tp]
\centering
\subfigure{\includegraphics[scale=0.5,clip=true,trim=0cm 0cm 0cm 0cm,]{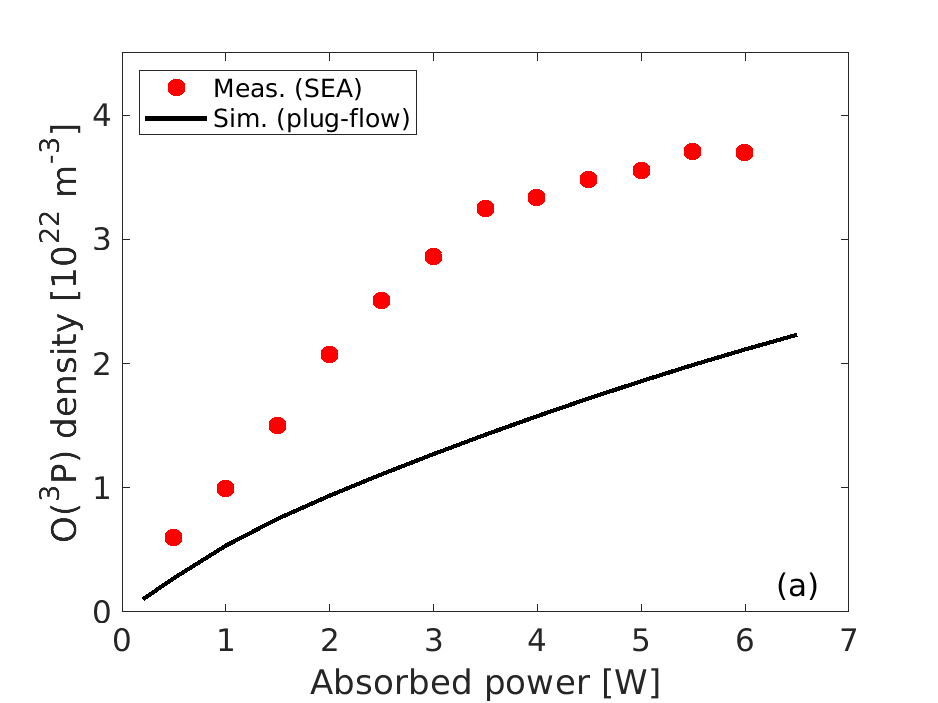}} \\ 
\subfigure{\includegraphics[scale=0.5,clip=true,trim=0cm 0cm 0cm 0cm,]{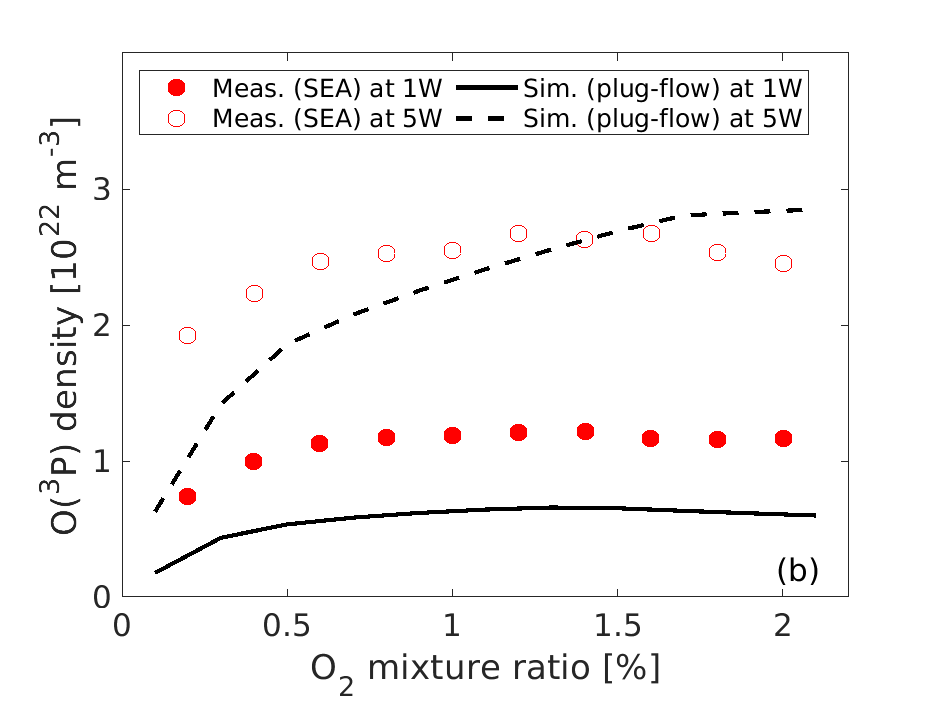}} \\ 
\subfigure{\includegraphics[scale=0.5,clip=true,trim=0cm 0cm 0cm 0cm,]{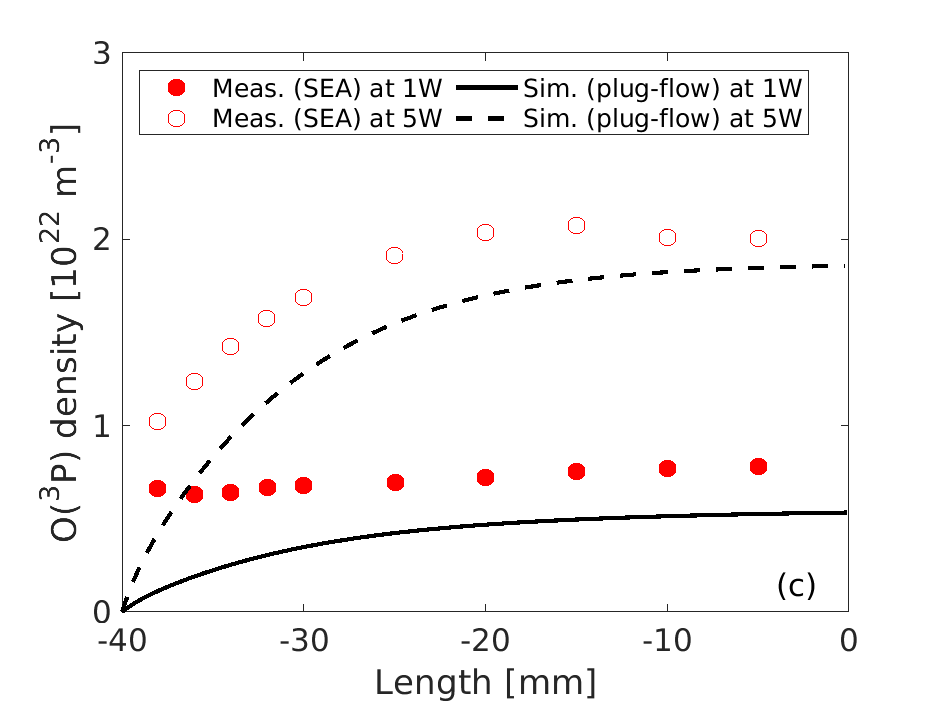}} 
\caption{
The $\mathrm{O(^3P)}$ densities of the He/$\mathrm{O_2}$ $\mu$APPJ reported by Winzer $et$ $al$ 2022 \cite{Win22}. (a) The SEA measurement data for a variation of the absorbed power from 0.5 W to 6.5 W (\tikzcircle[red, fill=red]{2.0pt}) and the corresponding plug-flow model calculations of this work (\textcolor{black}{\full}). (b) The SEA measurement data for a variation of the O$_2$ mixture ratio from $0.1\%$ to $2.0\%$ (\tikzcircle[red, fill=red]{2.0pt} and \tikzcircle[red, fill=white]{2.1pt} at 1 W and 5 W absorbed power, respectively) and the corresponding plug-flow model calculations of this work (\textcolor{black}{\full} and \textcolor{black}{\dashed}). The plug-flow model calculation values shown in figures (a) and (b) are the simulation results at the plasma channel exit, see section \ref{sec:mod}. (c) The one dimensional spatially resolved SEA measurement data along the gas flow direction in the plasma channel region (\tikzcircle[red, fill=red]{2.0pt} and \tikzcircle[red, fill=white]{2.1pt} at 1 W and 5 W absorbed power, respectively) and the corresponding plug-flow model calculations of this work (\textcolor{black}{\full} and \textcolor{black}{\dashed}). $1000$ sccm He gas flow mixed with $0.5\%$ O$_2$ (if not stated otherwise) are fed to the $1 \times 1 \times 40$ mm$^3$ plasma channel. 
}\label{fig:Win22}
\end{figure}

The $\mathrm{O(^3P)}$ densities of a He/$\mathrm{O_2}$ $\mu$APPJ (similar to the COST-Jet, but with a dielectric capillary between the electrodes) measured with the SEA approach by Winzer $et$ $al$ 2022 \cite{Win22} together with the corresponding plug-flow model calculations of this work are shown in figure \ref{fig:Win22}. The operating conditions are provided in the figure caption, and more details such as the measurement method and position are described in section \ref{sec:setup}. 
The simulated $\mathrm{O(^3P)}$ densities as a function of the absorbed power, $\mathrm{O_2}$ mixture ratio and plasma channel position in figure \ref{fig:Win22} slightly underestimate the corresponding SEA measurement data. The similar underestimation is also observed in figure \ref{fig:Wes16_Rie20_Ste22}(c). 
However, our simulation results still capture the overall trend and quantity of the measurement data. 
The measurement data in figures \ref{fig:Win22}(a) and \ref{fig:Win22}(b) are usually obtained at the middle or exit of the plasma channel, where both positions report the similar $\mathrm{O(^3P)}$ density under the considered operating conditions in this work (see figures \ref{fig:Ste21} and \ref{fig:Win22}(c)). The plug-flow model calculation values in figures \ref{fig:Win22}(a) and \ref{fig:Win22}(b) are the simulation results at the plasma channel exit, see section \ref{sec:mod}.
In figure \ref{fig:Win22}(a), the simulated and measured $\mathrm{O(^3P)}$ densities increase with increasing absorbed power, and this is similarly predicted in figure \ref{fig:Wes16_Rie20_Ste22}. The larger deviation between the measurements and simulations at larger absorbed power may be ascribed to experimental variations, since for instance the simulated $\mathrm{O(^3P)}$ density at 5 W in figure \ref{fig:Win22}(a) (around $1.9 \times 10^{22} \: \mathrm{m}^{-3}$) agrees better with the measurement data under the same operating conditions in figures \ref{fig:Win22}(b) (around $2.2 \times 10^{22} \: \mathrm{m}^{-3}$) and \ref{fig:Win22}(c) (around $2.0 \times 10^{22} \: \mathrm{m}^{-3}$) compared to those in figure \ref{fig:Win22}(a) (around $3.6 \times 10^{22} \: \mathrm{m}^{-3}$). 
It is indicated in figure \ref{fig:Win22}(b) that both the measured and simulated $\mathrm{O(^3P)}$ densities at 1 W and 5 W increase with increasing $\mathrm{O_2}$ mixture ratio, and start to saturate at $1.0\:\%$ $\mathrm{O_2}$ (similarly observed in figure \ref{fig:Mye21}).
In figure \ref{fig:Win22}(c), the measured and simulated $\mathrm{O(^3P)}$ densities at 1 W and 5 W increase continuously along the gas flow direction, and start to saturate at the middle of the plasma channel.

\subsection{Prediction accuracy of the simulated atomic oxygen densities relative to the measured ones}
\label{sec:statistik}

\begin{figure}[h!]
\centering
\includegraphics[scale=0.45,clip=true,trim=0cm 0cm 0cm 0cm,]{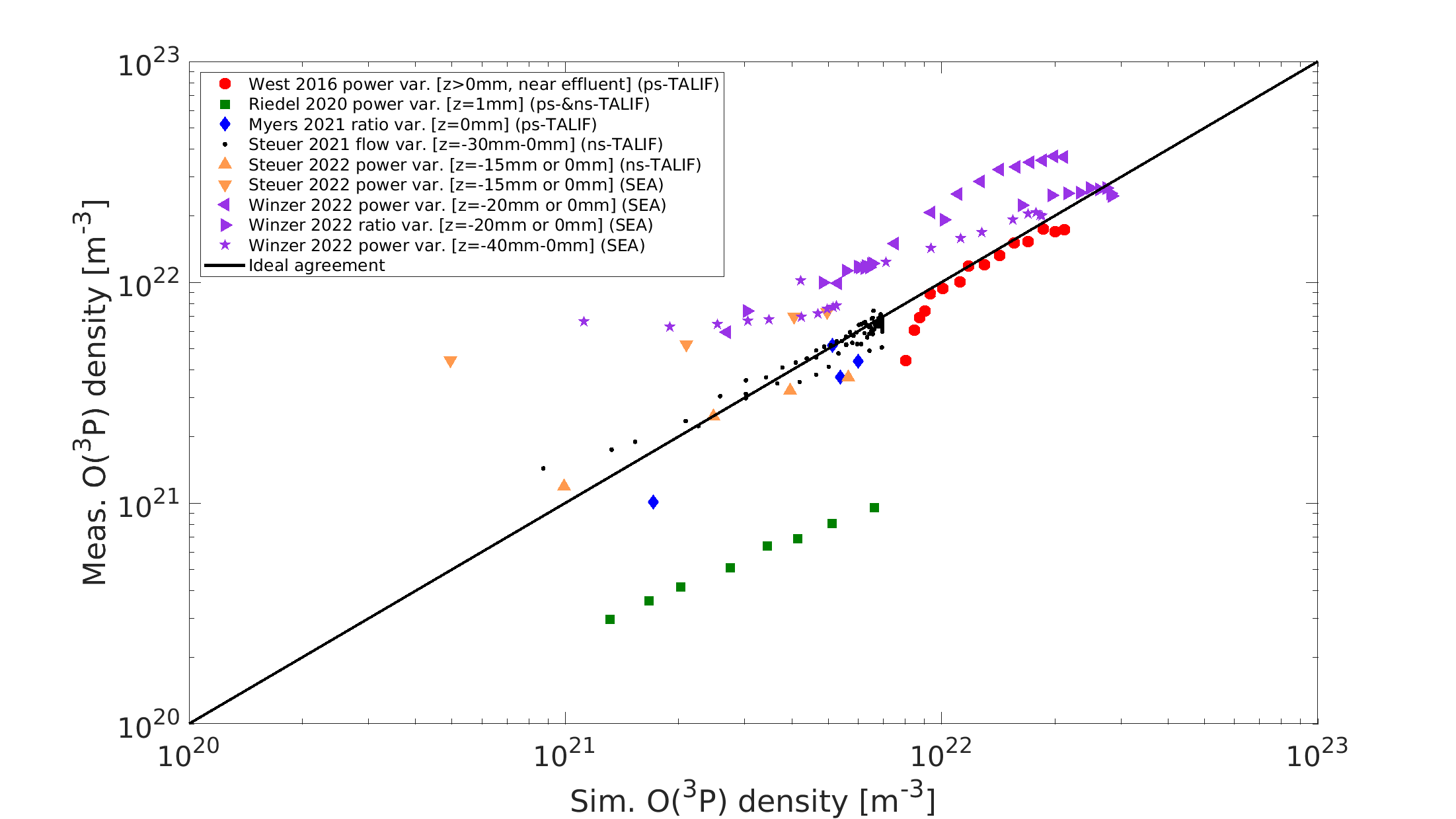}
\caption{
The measured atomic oxygen densities of several $\mu$APPJs from multiple publications \cite{Wes16,Rie20, Mye21, Ste21, Ste22, Win22} using the TALIF and the SEA measurement methods versus the corresponding simulated ones of this work using the plug-flow model. 
These density values are a summary of the validation data in section \ref{sec:validation}.
The measurement data as a function of the absorbed power (``power var."), the He gas flow rate (``flow var.") and the O$_2$ mixture ratio (``ratio var.") were reported by West 2016 \cite{Wes16}, Riedel $et$ $al$ 2020 \cite{Rie20}, Myers $et$ $al$ 2021 \cite{Mye21}, Steuer $et$ $al$ 2021 \cite{Ste21}, Steuer $et$ $al$ 2022 \cite{Ste22} and Winzer $et$ $al$ 2022 \cite{Win22}. The measurement data were collected at different positions, i.e. along the gas flow direction in the plasma channel region (z$<$0mm) \cite{Ste21,Win22}, at the middle or exit of the plasma channel region (z$\leqslant$0mm) \cite{Mye21,Ste22,Win22}, and in the near effluent region (z$>$0mm) \cite{Wes16,Rie20}.
For the sake of consistency, only the plug-flow model calculation results at the plasma channel exit (z$=$0mm) are used to compared with the measurement data at the middle or exit of the plasma channel region (z$\leqslant$0mm) \cite{Mye21,Ste22,Win22} and those in the near effluent region (z$>$0mm) \cite{Wes16,Rie20}, see section \ref{sec:mod}.
The operating conditions of the measurements are the same as those of the corresponding simulations.
Therefore, an ideal agreement between the measurement data and simulation results is illustrated by the black solid line, where the measured densities are equal to the simulated densities under the same operating conditions. 
}\label{fig:statistic_1_summary}
\end{figure}

A summary of the validation data of the multiple $\mu$APPJs in section \ref{sec:validation} is presented with the measured atomic oxygen densities \cite{Wes16,Rie20, Mye21, Ste21, Ste22, Win22} versus the corresponding simulated ones of this work in figure \ref{fig:statistic_1_summary}. 
The key information are summarized as follows.
The measurement data were collected at different positions: i.e. along the gas flow direction in the plasma channel region \cite{Ste21,Win22}, at the middle or exit of the plasma channel region \cite{Mye21,Ste22,Win22}, and in the near effluent region \cite{Wes16,Rie20}. 
For the sake of consistency, only the simulation results at the plasma channel exit are used to compared with the measurement data, which were collected at the middle or exit of the plasma channel region \cite{Mye21,Ste22,Win22} and collected in the near effluent region \cite{Wes16,Rie20}, see section \ref{sec:mod}. 
The plug-flow model calculation results along the gas flow direction in the plasma channel region are used to compared with the corresponding one dimensional spatially resolved measurement data \cite{Ste21,Win22}.
The operating conditions of the measurements are the same as those of the corresponding simulations.

It should be emphasized that an ideal agreement between the measured and simulated atomic oxygen densities is illustrated by the black solid line shown in figure \ref{fig:statistic_1_summary}, where the measured densities are equal to the simulated densities under the same operating conditions. 
However, the small deviation between the measured and simulated atomic oxygen densities is realistically inevitable, e.g. the slight influence due to model limitation, experimental error, and potential inconsistency between the input parameters of the simulations and measurements. 
Therefore, points near the aforementioned black solid line can also be regarded as a good agreement.
Most values of the measurement data and simulation results in figure \ref{fig:statistic_1_summary} are near the black solid line, and evenly distributed on both sides. Specifically, a good agreement is observed between the simulations and most of the TALIF measurements with a variation of the absorbed power, the He gas flow rate and the O$_2$ mixture ratio. The SEA measurement data are overall slightly larger than the simulation results.

One focus of this work is dedicated to quantifying the prediction accuracy of our simulated atomic oxygen densities relative to the measured ones, and further analyzing the influence of the absence of the dominant atomic oxygen gain and loss reaction channels on the aforementioned prediction accuracy. 
The prediction accuracy, as a function of the deviation between the measurement data and simulation results, is essentially influenced by factors such as model limitation, experimental error, and potential inconsistency between the input parameters of the simulations and measurements. 
It is visibly observed in figure \ref{fig:statistic_1_summary} that the SEA measurement data overall slightly overestimate the simulation results. 
Furthermore, a slight overestimation of the SEA measurement data relative to the TALIF ones was recently reported by Steuer $et$ $al$ \cite{Ste22}. 
In view of these, the SEA measurement data are excluded in the following investigation of the model prediction accuracy. Therefore, we focus on the prediction accuracy of the simulations relative to the TALIF measurements.

\begin{figure}[h!]
\begin{minipage}[b]{0.5\textwidth}
\includegraphics[width=0.95\textwidth]{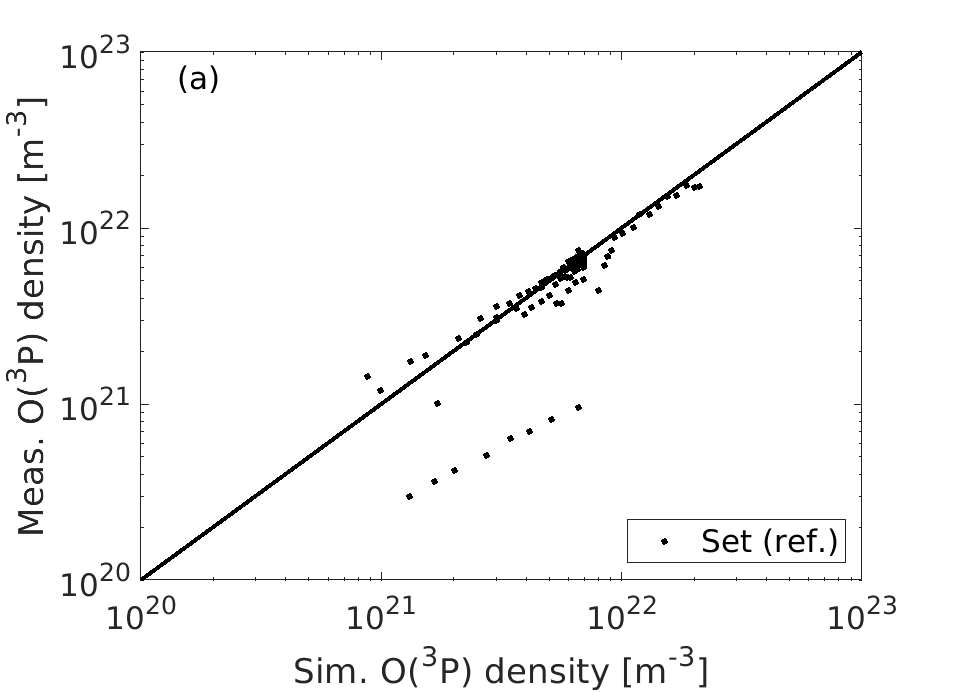} \\
\end{minipage}
\begin{minipage}[b]{0.5\textwidth}
\includegraphics[width=0.95\textwidth]{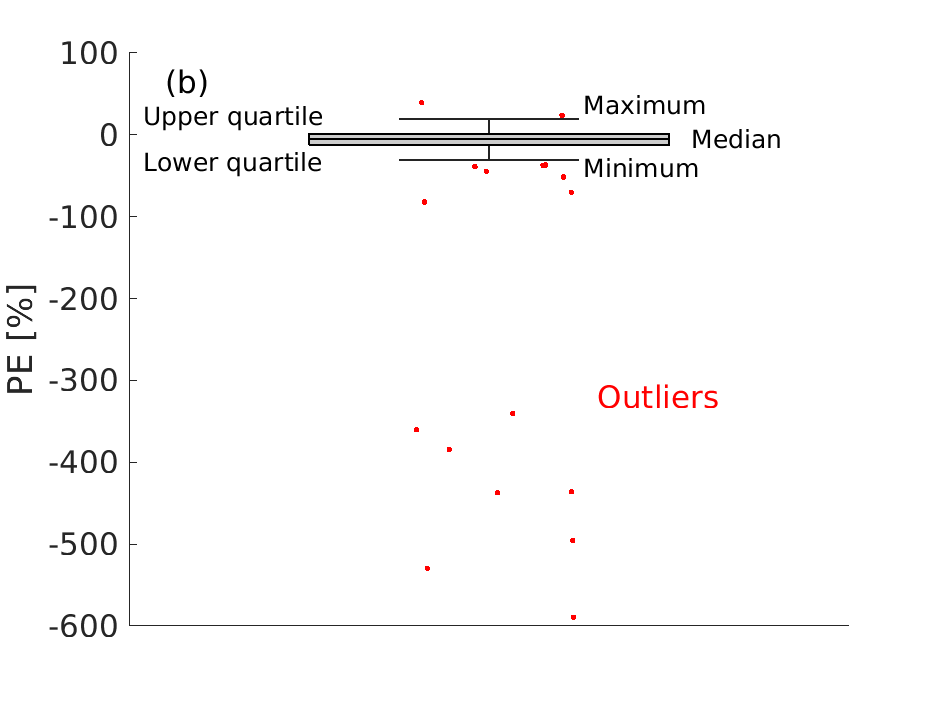} \\ 
\end{minipage}
\caption{
(a) The measured atomic oxygen densities of several $\mu$APPJs from multiple publications \cite{Wes16,Rie20, Mye21, Ste21, Ste22} using the TALIF measurement method versus the corresponding simulated ones of this work using the plug-flow model. The atomic oxygen densities are identical to those in figure \ref{fig:statistic_1_summary} excluding the SEA measurement data. The densities are calculated with the reference chemical kinetics (i.e. the reaction ``Set (ref.)") reported in \ref{sec:chemkin}. 
(b) The box plot visualization of the percentage error of the measured and simulated atomic oxygen densities in figure \ref{fig:statistic_2_TALIF}(a). 
}\label{fig:statistic_2_TALIF}
\end{figure}

The majority of the data points using the TALIF in figure \ref{fig:statistic_1_summary} are close to the black solid line, while some of the data points using the TALIF are further from the black solid line than the majority likely due to multiple factors. In the case that these distant data points can be considered as outliers, we choose to exclude them for a reasonable prediction accuracy calculation, as they are not representative of the majority of the measurement data. 
The measured atomic oxygen densities using the TALIF \cite{Wes16,Rie20, Mye21, Ste21, Ste22} versus the corresponding simulated ones using the plug-flow model of this work are presented in figure \ref{fig:statistic_2_TALIF}(a). The atomic oxygen densities in figure \ref{fig:statistic_2_TALIF}(a) are identical to those in figure \ref{fig:statistic_1_summary} excluding the SEA measurement data. These densities are calculated with the reference chemical kinetics (i.e. the reaction ``Set (ref.)") reported in \ref{sec:chemkin}. 
The prediction accuracy of our simulation results relative to the aforementioned TALIF measurement data is quantified by the percentage error straightforwardly providing the degree of underprediction and overprediction \cite{Vis21}. The percentage error is given by
\begin{equation}
{\rm percentage \: error \: (PE)} = \frac{n_{meas.}(i)-n_{sim.}(i)}{n_{meas.}(i)}  \cdot  100\%,
\label{eqn:PE}
\end{equation}
where $n_{meas.}(i)$ and $n_{sim.}(i)$ are the measured and simulated atomic oxygen densities of a certain data point $i$, respectively. 
A box plot of the percentage error of the measured and simulated atomic oxygen densities in figure \ref{fig:statistic_2_TALIF}(a) is visualized in figure \ref{fig:statistic_2_TALIF}(b). 
The box plot as a simple and straightforward detection technique \cite{Hod04,Wan19,Smi20,Sik23} provides a boundary to visually pinpoint outliers. 
The boundary is defined at a minimum and a maximum which are away from the lower quartile and the upper quartile for 1.5$\times$distance between the lower and upper quartiles, respectively \cite{Hod04,Smi20}, where the lower quartile is the 25th percentile and the upper quartile is the 75th percentile. 
The data points outside the minimum and maximum are regarded as the outliers.
It is found in figure \ref{fig:statistic_2_TALIF}(b) that eight data points located in the range of $-600\% - -300\%$ are significantly far away from the minimum. Therefore, these eight data points are treated as significant outliers, and excluded in the following investigation of the model prediction accuracy relative to the TALIF measurements. On the other hand, there are also some other data points outside the minimum and maximum, which are also defined as outliers by the box plot method. However these data points are much closer to the minimum and maximum compared to the aforementioned eight data points. Furthermore, these data points are part of the measurement data sets from specific publications, where most of the measurement data are not defined as outliers. In view of this, these data points are still included in the following investigation of the model prediction accuracy relative to the TALIF measurements, even though they have been identified as outliers.

\begin{figure}[h!]
\begin{minipage}[b]{0.5\textwidth}
\includegraphics[width=0.95\textwidth]{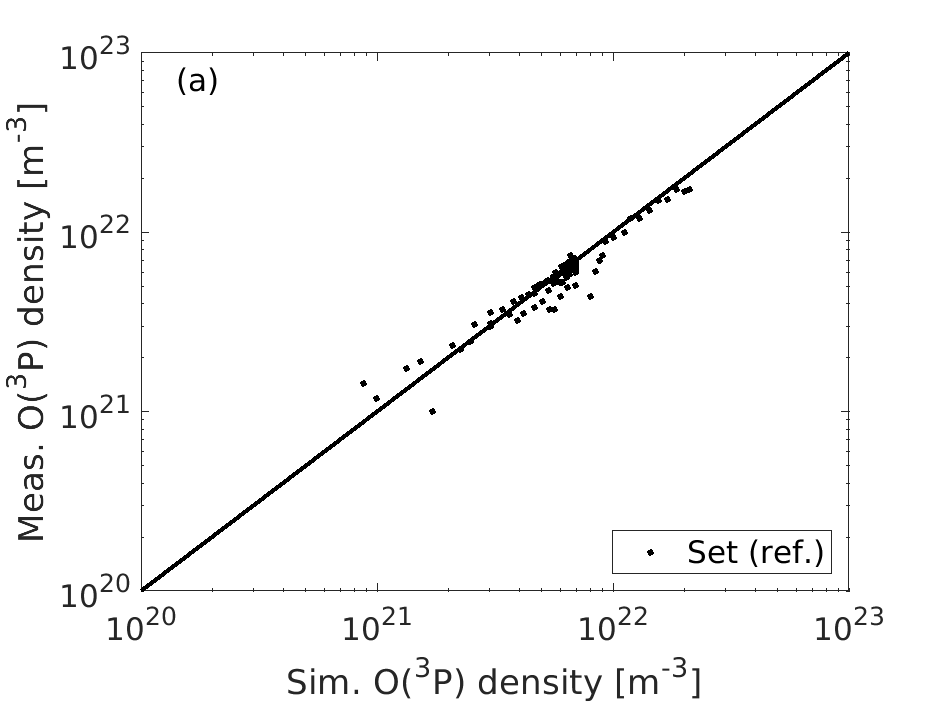} \\
\end{minipage}
\begin{minipage}[b]{0.5\textwidth}
\includegraphics[width=0.95\textwidth]{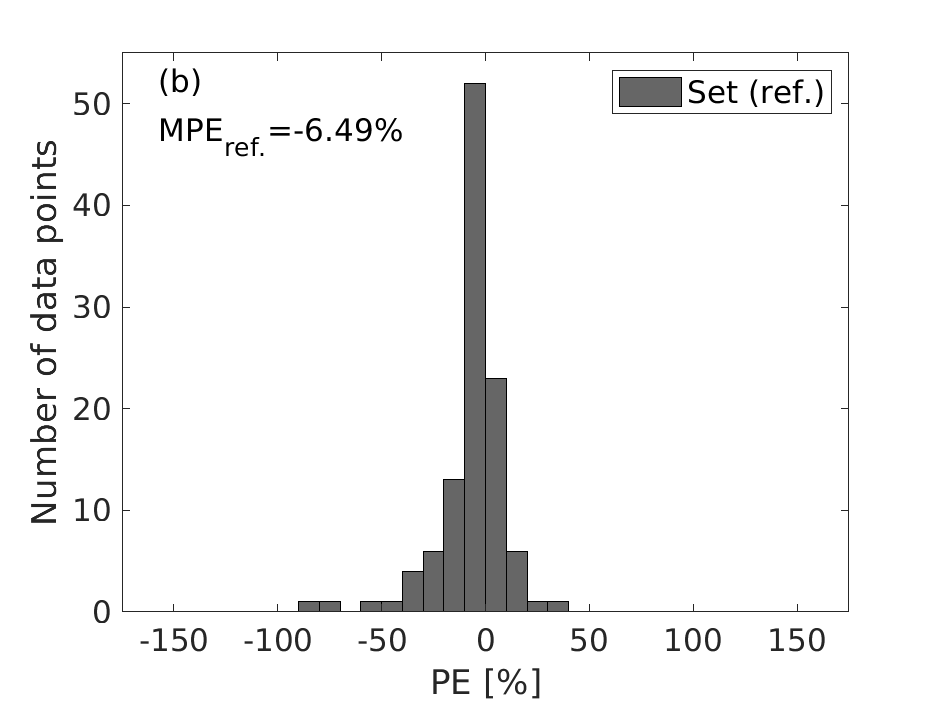} \\ 
\end{minipage}
\caption{
(a) The measured atomic oxygen densities of several $\mu$APPJs from multiple publications \cite{Wes16, Mye21, Ste21, Ste22} using the TALIF measurement method versus the corresponding simulated ones of this work using the plug-flow model. The atomic oxygen densities are identical to those in figure \ref{fig:statistic_2_TALIF}(a) excluding the eight significant outliers detected by the box plot in figure \ref{fig:statistic_2_TALIF}(b). The densities are calculated with the reference chemical kinetics (i.e. the reaction ``Set (ref.)") reported in \ref{sec:chemkin}. 
(b) The histogram visualization of the number of data points versus the percentage error of the measured and simulated atomic oxygen densities in figure \ref{fig:statistic_3_TALIF_noRie20}(a). 
}\label{fig:statistic_3_TALIF_noRie20}
\end{figure}

The atomic oxygen densities \cite{Wes16, Mye21, Ste21, Ste22} in figure \ref{fig:statistic_3_TALIF_noRie20}(a) are identical to those in figure \ref{fig:statistic_2_TALIF}(a) excluding the eight significant outliers detected by the box plot in figure \ref{fig:statistic_2_TALIF}(b). These densities are calculated with the reference chemical kinetics (i.e. the reaction ``Set (ref.)") reported in \ref{sec:chemkin}. 
An intuitive distribution of the percentage error (i.e. the degree of underprediction and overprediction) can be given by a histogram plot. Therefore, the histogram plot is visualized in figure \ref{fig:statistic_3_TALIF_noRie20}(b) with the number of the data points versus the percentage error of the measured and simulated atomic oxygen densities in figure \ref{fig:statistic_3_TALIF_noRie20}(a). 
As expected, an approximate normal distribution of the percentage error is observed in the histogram plot of figure \ref{fig:statistic_3_TALIF_noRie20}(b), and most of the values are close to $0\%$. 
In addition, the mean percentage error \cite{Vis21} is used in this work to provide one single value describing the model prediction accuracy relative to all the measurement data \cite{Wes16, Mye21, Ste21, Ste22}. 
The mean percentage error is calculated by
\begin{equation}
{\rm mean \: percentage \: error \: (MPE)} = \frac{1}{L} \sum_{i=1}^L \frac{n_{meas.}(i)-n_{sim.}(i)}{n_{meas.}(i)}  \cdot  100\%,
\label{eqn:MPE}
\end{equation}
where $L$ is the number of the data points.
The mean percentage error in the case of the simulations using the reaction ``Set (ref.)" is MPE$_{ref.}=-6.49\%$, close to $0\%$. This quantitatively indicates that the atomic oxygen densities of the $\mu$APPJs measured by various research groups \cite{Wes16, Mye21, Ste21, Ste22} with the TALIF method can be well predicted by the model of this work.

\begin{table}[h!]\scriptsize
\setlength{\tabcolsep}{14pt}
\centering
\caption{
The first three dominant gain and loss reaction channels of the atomic oxygen species $\mathrm{O(^3 P)}$ and their corresponding contribution at the plasma channel exit of the COST-Jet under a typical operating condition, i.e. 1000 sccm He mixed with $0.5\%$ O$_2$ is fed into the plasma channel driven by the absorbed power of 1.00 W. ``O$_2$" in this table represents O$_2(v=0)$. The results are obtained from the plug-flow model calculations using the reference chemical kinetics (i.e. the reaction ``Set (ref.)") reported in \ref{sec:chemkin}.
}
\begin{tabular}{lllll}
\\[\dimexpr-\normalbaselineskip+3pt]
\hline
\\[\dimexpr-\normalbaselineskip+3pt]
\# &Reaction     &  Contribution  &  Table  &  \\ \hline
\\[\dimexpr-\normalbaselineskip+3pt]
& Dominant gain reaction channels of $\mathrm{O(^3 P)}$ &   &    & \\ 
\\[\dimexpr-\normalbaselineskip+3pt]
1  & $ e + \mathrm{O_2} \rightarrow e + 2 \mathrm{O(^3 P)}  $ & 28.74$\%$ &     \ref{tab:ReactionListHeO2}(R9) &    \\ 
\\[\dimexpr-\normalbaselineskip+3pt]
2  & $ \mathrm{O_2 (b^1 \Sigma_{g^+})} + \mathrm{O_3} \rightarrow 2 \mathrm{O_2} + \mathrm{O(^3 P)}  $ & 20.71$\%$ & \ref{tab:ReactionListHeO2updated}(R11) &    \\ 
\\[\dimexpr-\normalbaselineskip+3pt]
3  & $ e + \mathrm{O_2} \rightarrow e + \mathrm{O(^1 D)} + \mathrm{O(^3 P)}  $ & 14.40$\%$  &  \ref{tab:ReactionListHeO2}(R10)  & \\ 
\\[\dimexpr-\normalbaselineskip+6pt]
& Dominant loss reaction channels of $\mathrm{O(^3 P)}$  &   &    & \\
\\[\dimexpr-\normalbaselineskip+3pt]
4  & $ \mathrm{O(^3P)} + \mathrm{wall} \rightarrow 1/2\mathrm{O}_2  $ & 34.83$\%$ &  \ref{tab:wrHeO2}(R14)  &   \\ 
\\[\dimexpr-\normalbaselineskip+3pt]
5  & $ \mathrm{He} + \mathrm{O(^3 P)} + \mathrm{O_2} \rightarrow \mathrm{He} + \mathrm{O_3}  $ & 26.24$\%$  & \ref{tab:ReactionListHeO2}(R95)  &  \\ 
\\[\dimexpr-\normalbaselineskip+3pt]
6  & $ \mathrm{He} + 2 \mathrm{O(^3 P)} \rightarrow \mathrm{He} + \mathrm{O_2 (b^1 \Sigma_{g^+})}  $ & 13.03$\%$  & \ref{tab:ReactionListHeO2updated}(R86) &  \\ 
\\[\dimexpr-\normalbaselineskip+3pt]

\\[\dimexpr-\normalbaselineskip+3pt]
\hline
\end{tabular}

\label{tab:reacDomi}
\end{table}

The first three dominant atomic oxygen gain and loss reaction channels and their corresponding contribution at the plasma channel exit of the COST-Jet are reported in table \ref{tab:reacDomi}. 
As noted earlier, one focus of this work is dedicated to analyzing the influence of an absence of the dominant atomic oxygen gain and loss reaction channel on our model prediction accuracy relative to the measurements from various research groups.
Specifically, the influence of removing each of these dominant reaction channels on the model prediction accuracy is investigated in the following texts.
Note that table \ref{tab:reacDomi} is only valid for the COST-Jet under a typical operating condition, i.e. 1000 sccm He mixed with 0.5$\%$ O$_2$ is fed into the plasma channel driven by the absorbed power of 1.00 W. 
It should be emphasized that the dominant reaction channels and the corresponding contribution vary at different plasma channel positions of the $\mu$APPJs and change with a variation of the operating conditions such as the absorbed power, the He gas flow rate and the O$_2$ mixture ratio.
An analysis of the full picture of these dominant reaction channels at each plasma channel position and each operating condition deviates from our main focus: i.e. the model prediction accuracy mentioned above. For the sake of the simplicity, only the dominant reaction channels at the plasma channel exit of the COST-Jet under the typical operating condition in table \ref{tab:reacDomi} are analyzed with our plug-flow model calculations using the reference chemical kinetics (i.e. the reaction ``Set (ref.)") reported in \ref{sec:chemkin}.

\begin{figure}[tp]
\begin{minipage}[b]{0.5\textwidth}
\includegraphics[width=0.95\textwidth]{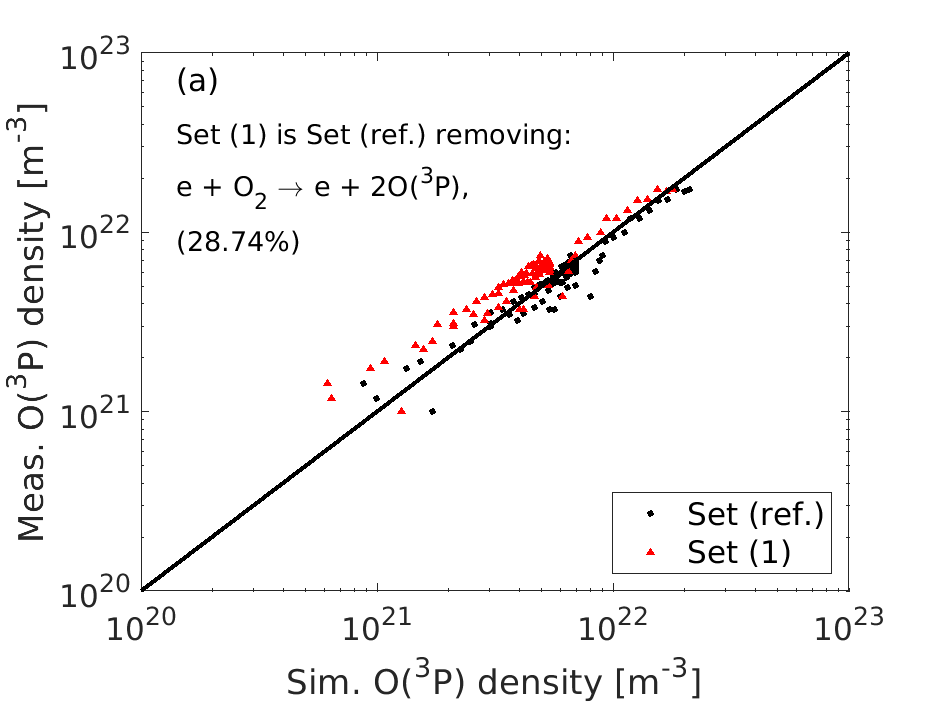} \\
\end{minipage}
\begin{minipage}[b]{0.5\textwidth}
\includegraphics[width=0.95\textwidth]{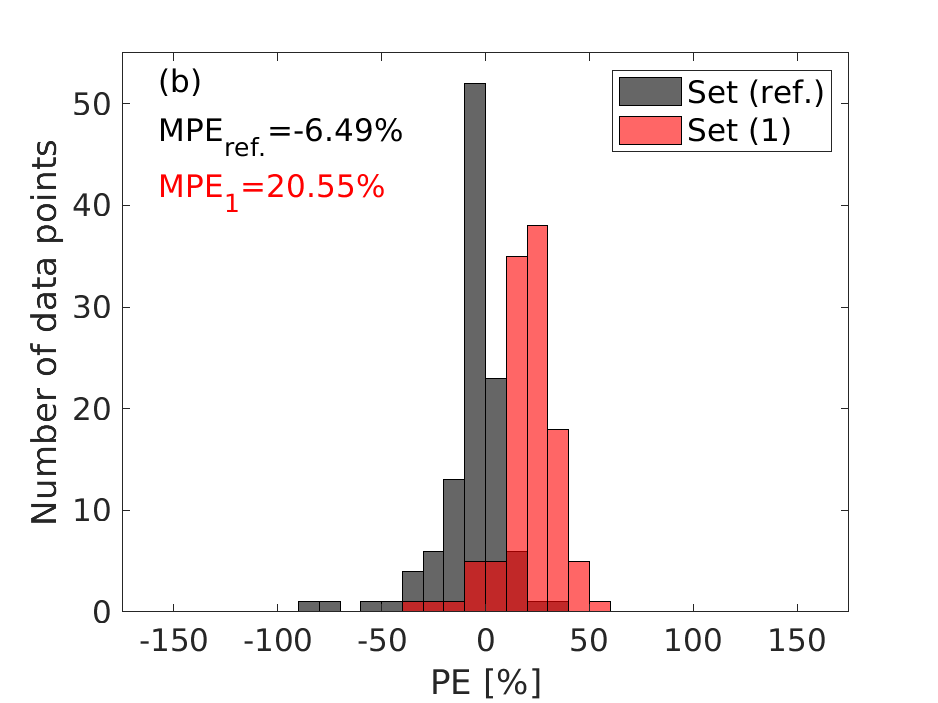} \\ 
\end{minipage}
\begin{minipage}[b]{0.5\textwidth}
\includegraphics[width=0.95\textwidth]{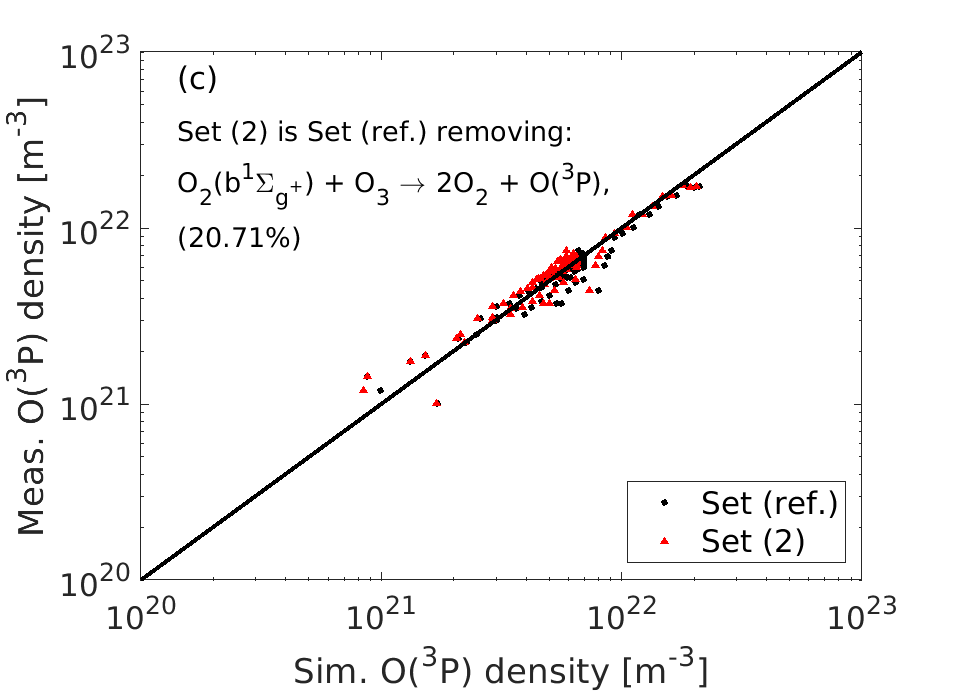} \\ 
\end{minipage}
\begin{minipage}[b]{0.5\textwidth}
\includegraphics[width=0.95\textwidth]{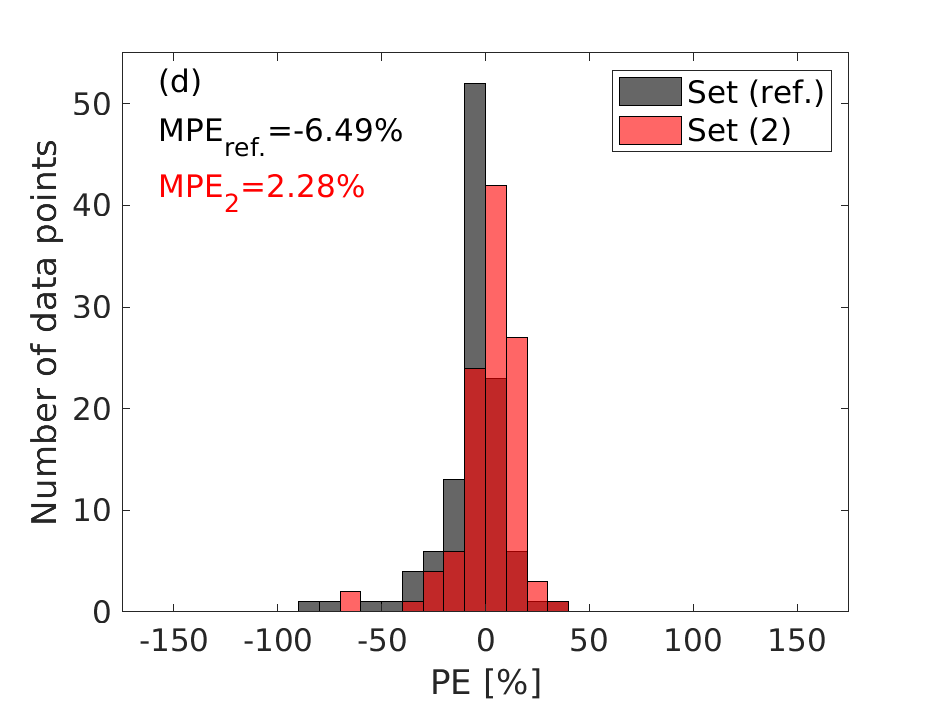} \\ 
\end{minipage}
\begin{minipage}[b]{0.5\textwidth}
\includegraphics[width=0.95\textwidth]{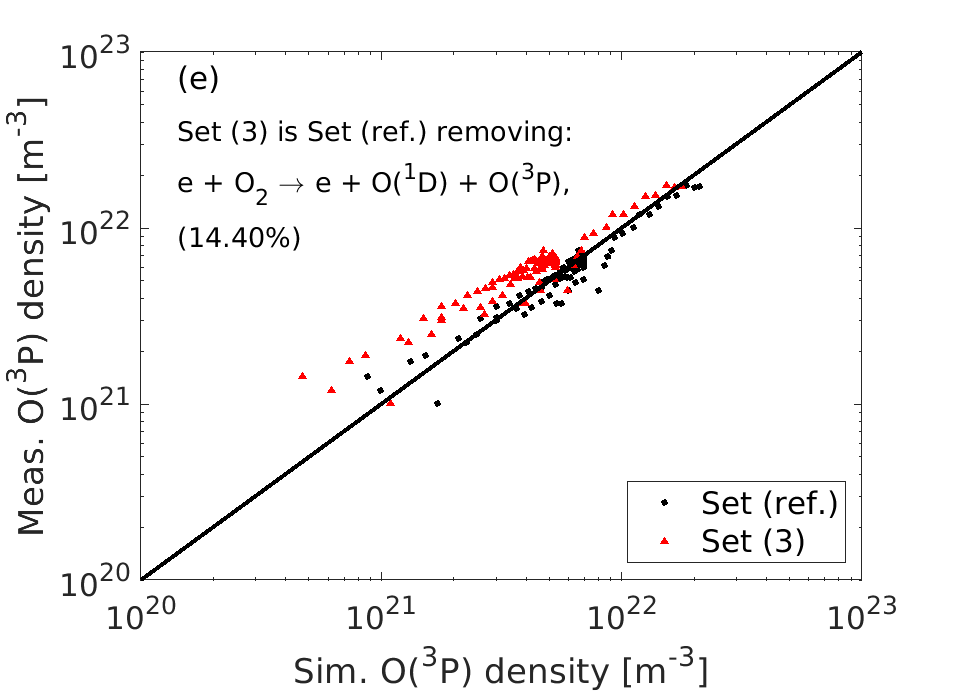} \\ 
\end{minipage}
\begin{minipage}[b]{0.5\textwidth}
\includegraphics[width=0.95\textwidth]{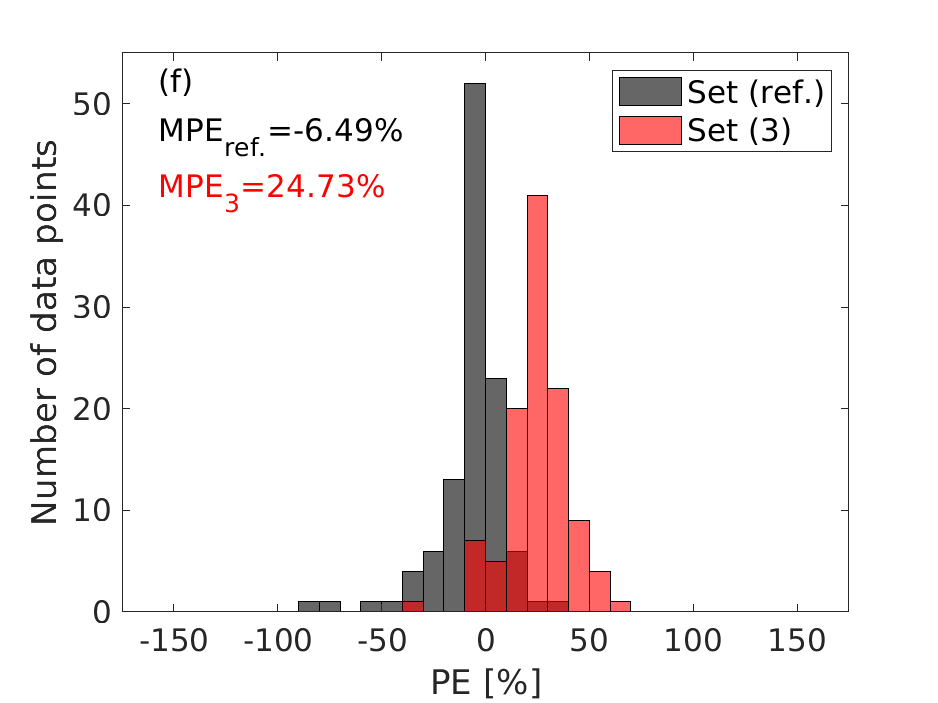} \\ 
\end{minipage}
\caption{
The influence of removing each of the dominant atomic oxygen gain reaction channels in table \ref{tab:reacDomi} in the simulations on our model prediction accuracy relative to the TALIF measurements from multiple publications \cite{Wes16, Mye21, Ste21, Ste22}. The measured atomic oxygen densities versus the corresponding simulated ones and the histogram visualization of the corresponding percentage error using the reaction ``Set (1)", ``Set (2)" or ``Set (3)" in the simulations are presented in a same way as those using the reaction ``Set (ref.)" in the simulations in figures \ref{fig:statistic_3_TALIF_noRie20}(a) and \ref{fig:statistic_3_TALIF_noRie20}(b), respectively. 
The reaction ``Set (ref.)" are the reference chemical kinetics reported in \ref{sec:chemkin}. The reactions ``Set (1)" in figures (a)-(b), ``Set (2)" in figures (c)-(d) and ``Set (3)" in figures (e)-(f) are the reaction ``Set (ref.)" removing reactions (1), (2) and (3) in table \ref{tab:reacDomi}, respectively.
}\label{fig:statistic_4_sou}
\end{figure}

The influence of removing each of the dominant atomic oxygen gain reaction channels in table \ref{tab:reacDomi} in the simulations on our model prediction accuracy are presented in figure \ref{fig:statistic_4_sou}.
It is found in figures \ref{fig:statistic_4_sou}(a), \ref{fig:statistic_4_sou}(c) and \ref{fig:statistic_4_sou}(e) that most of the atomic oxygen density data points in the case of using the reaction ``Set (ref.)" are evenly distributed on both sides of the black solid line, while those using the reactions ``Set (1)", ``Set (2)" and ``Set (3)" are above the black solid line. The influence of the dominant gain reaction channels on the model prediction accuracy is illustrated more clearly in figures \ref{fig:statistic_4_sou}(b), \ref{fig:statistic_4_sou}(d) and \ref{fig:statistic_4_sou}(f) that the percentage error histogram plot in the case of using the reaction ``Set (ref.)" is approximately a normal distribution with a mean of $-6.49\%$, while the means of the approximate normal distributions in the case of using the reactions ``Set (1)", ``Set (2)" and ``Set (3)" are shifted to the right and to be positive values of $20.55\%$, $2.28\%$ and $24.73\%$, respectively.
In other words, the simulation results in the case of removing one of the dominant atomic oxygen gain reaction channels overall underestimate the measurement data in figure \ref{fig:statistic_4_sou}. This is ascribed to the lower production rate of atomic oxygen in the case of the absence of a corresponding dominant gain reaction channel in the simulations.
Moreover, it is observed in figure \ref{fig:statistic_4_sou} that the degree to which the mean percentage error is shifted is not directly correlated with the contribution of the corresponding dominant atomic oxygen gain reaction channel. For instance, the second dominant reaction channel $ \mathrm{O_2 (b^1 \Sigma_{g^+})} + \mathrm{O_3} \rightarrow 2 \mathrm{O_2} + \mathrm{O(^3 P)} $ contributes  20.71$\%$ of atomic oxygen production. However, the mean percentage error is only shifted by $ \mathrm{MPE_2 - MPE_{ref.}} = 2.28\% - (-6.49\%) = 8.77\%$ in the simulations removing this dominant reaction channel. In contrast, the third dominant reaction channel $ e + \mathrm{O_2} \rightarrow e + \mathrm{O(^1 D)} + \mathrm{O(^3 P)} $ contributes 14.40$\%$ of atomic oxygen production, but the mean percentage error is even shifted by $ \mathrm{MPE_3 - MPE_{ref.}} = 24.73\% - (-6.49\%) = 31.22\%$ in the simulations excluding this dominant reaction.
It should be emphasized that this non-correlation is potentially caused by diverse factors, e.g. including but not limited to the following three points. (i) As noted earlier, the dominant reaction channels and especially the corresponding contribution vary at different plasma channel position of the $\mu$APPJs and change with a variation of the operating conditions such as the absorbed power, the He gas flow rate and the O$_2$ mixture ratio. (ii) The number of the corresponding data points at different plasma channel positions under different operating conditions involved in this work deviates from each other. (iii) The corresponding complex chemical kinetics defining the atomic oxygen density at certain plasma channel positions under certain operating conditions are affected by the absence of an associated dominant reaction channel.

\begin{figure}[tp]
\begin{minipage}[b]{0.5\textwidth}
\includegraphics[width=0.95\textwidth]{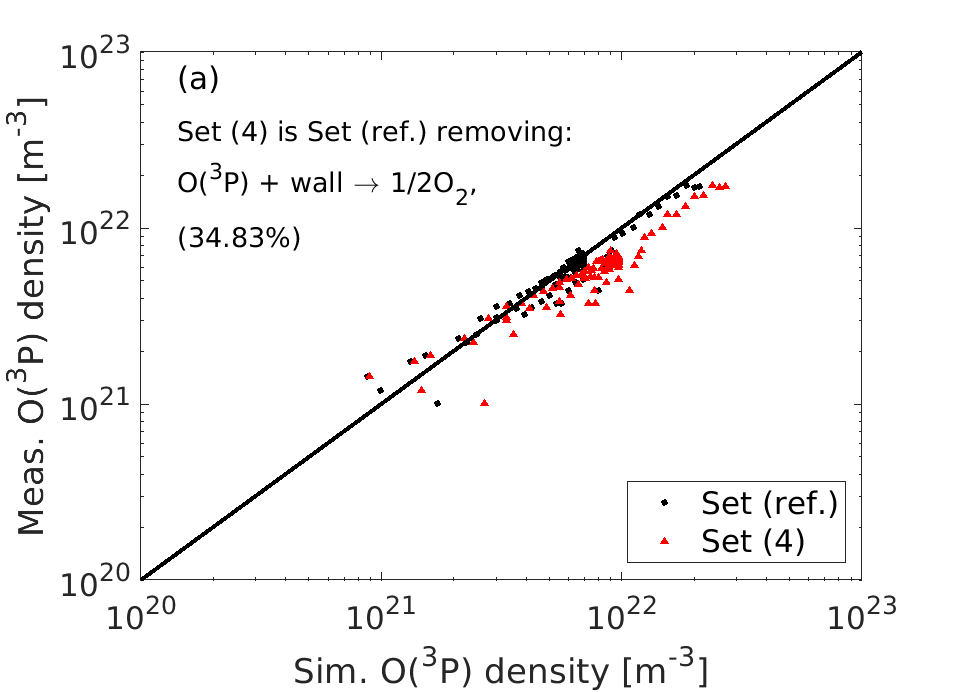} \\
\end{minipage}
\begin{minipage}[b]{0.5\textwidth}
\includegraphics[width=0.95\textwidth]{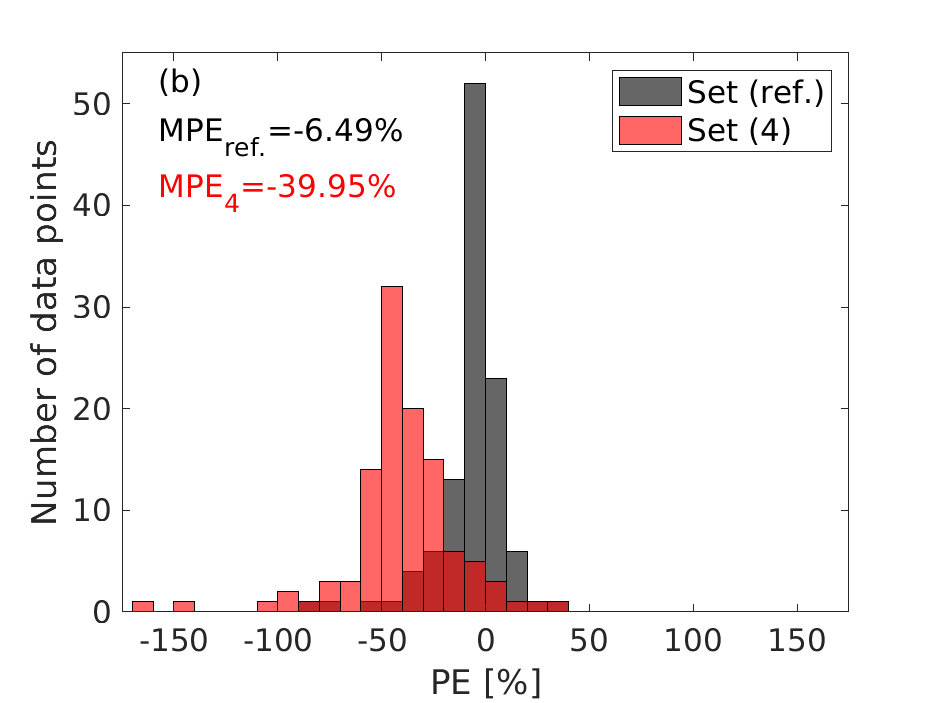} \\ 
\end{minipage}
\begin{minipage}[b]{0.5\textwidth}
\includegraphics[width=0.95\textwidth]{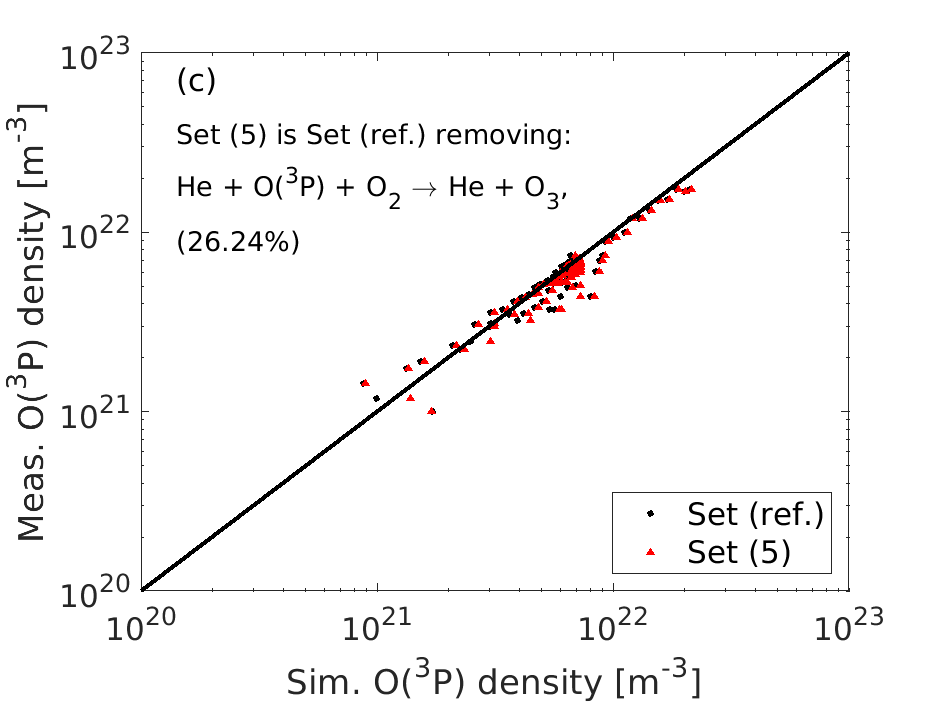} \\ 
\end{minipage}
\begin{minipage}[b]{0.5\textwidth}
\includegraphics[width=0.95\textwidth]{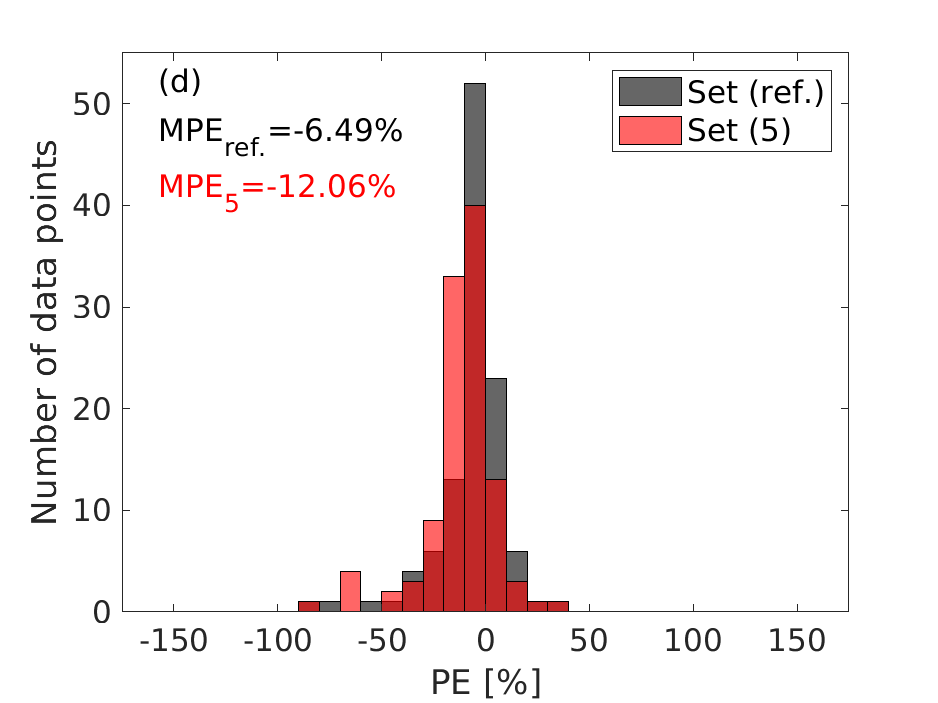} \\ 
\end{minipage}
\begin{minipage}[b]{0.5\textwidth}
\includegraphics[width=0.95\textwidth]{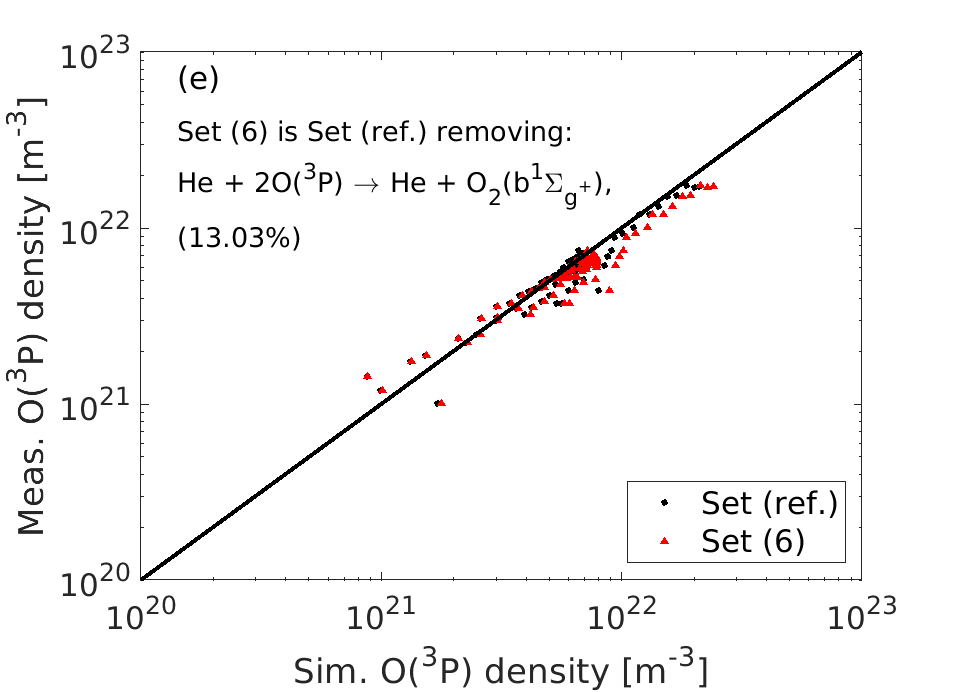} \\ 
\end{minipage}
\begin{minipage}[b]{0.5\textwidth}
\includegraphics[width=0.95\textwidth]{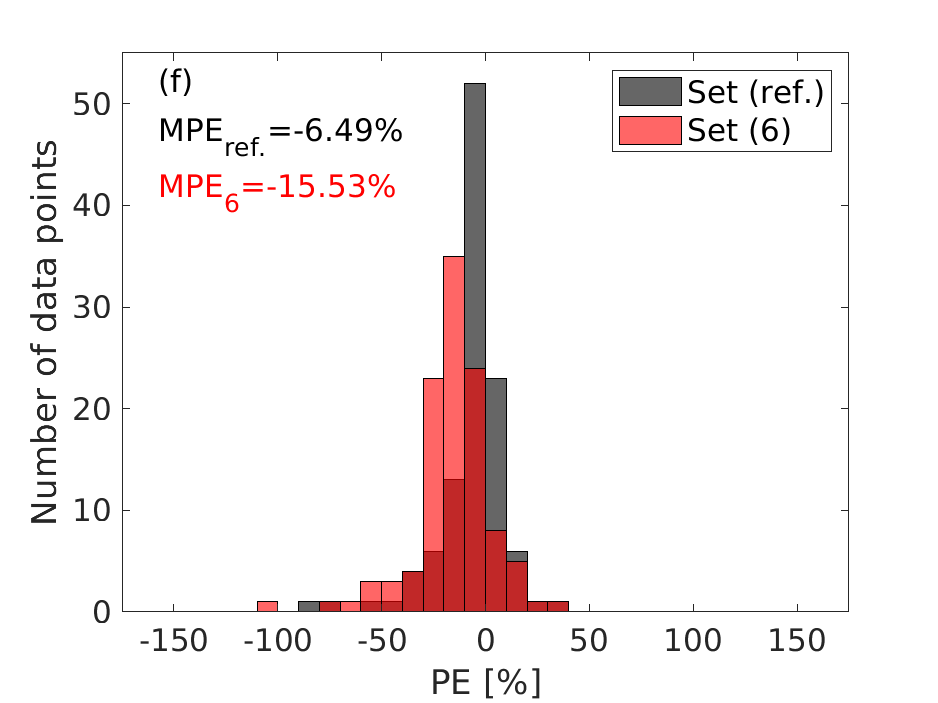} \\ 
\end{minipage}
\caption{
The influence of removing each of the dominant atomic oxygen loss reaction channels in table \ref{tab:reacDomi} in the simulations on our model prediction accuracy relative to the TALIF measurements from multiple publications \cite{Wes16, Mye21, Ste21, Ste22}. The measured atomic oxygen densities versus the corresponding simulated ones and the histogram visualization of the corresponding percentage error using the reaction ``Set (4)", ``Set (5)" or ``Set (6)" in the simulations are presented in a same way as those using the reaction ``Set (ref.)" in the simulations in figures \ref{fig:statistic_3_TALIF_noRie20}(a) and \ref{fig:statistic_3_TALIF_noRie20}(b), respectively. The reaction ``Set (ref.)" are the reference chemical kinetics reported in \ref{sec:chemkin}. 
The reactions ``Set (4)" in figures (a)-(b), ``Set (5)" in figures (c)-(d) and ``Set (6)" in figures (e)-(f) are the reaction ``Set (ref.)" removing reactions (4), (5) and (6) in table \ref{tab:reacDomi}, respectively.
}\label{fig:statistic_5_sin}
\end{figure}

The influence of removing each of the dominant atomic oxygen loss reaction channels in table \ref{tab:reacDomi} in the simulations on our model prediction accuracy are presented in figure \ref{fig:statistic_5_sin}. The above-mentioned outcomes in figure \ref{fig:statistic_4_sou} are similarly obtained in figure \ref{fig:statistic_5_sin}. The exception is that three reaction sets, i.e. ``Set (4)", ``Set (5)" and ``Set (6)", are obtained by removing reactions (4), (5) and (6) in table \ref{tab:reacDomi} in the reaction ``Set (ref.)", respectively. Most of the atomic oxygen density data points using the reactions ``Set (4)", ``Set (5)" and ``Set (6)" are under the black solid line, and the means of the approximate normal distributions of the corresponding percentage error are shifted to the left and to the negative values of $-39.95\%$, $-12.06\%$ and $-15.53\%$, respectively. In other words, the simulation results in the case of removing one of the dominant atomic oxygen loss reaction channels overall overestimate the measurement data in figure \ref{fig:statistic_5_sin} . This is owing to the lower destruction rate of atomic oxygen in the case of the absence of a corresponding dominant loss reaction channel in the simulations. Similar non-correlation is obtained between the shifted degree of the mean percentage error and the contribution of the corresponding dominant atomic oxygen loss reaction channel, due to the potential causes mentioned above.

\begin{figure}[tp]
\centering
\includegraphics[scale=0.45,clip=true,trim=0cm 0cm 0cm 0cm,]{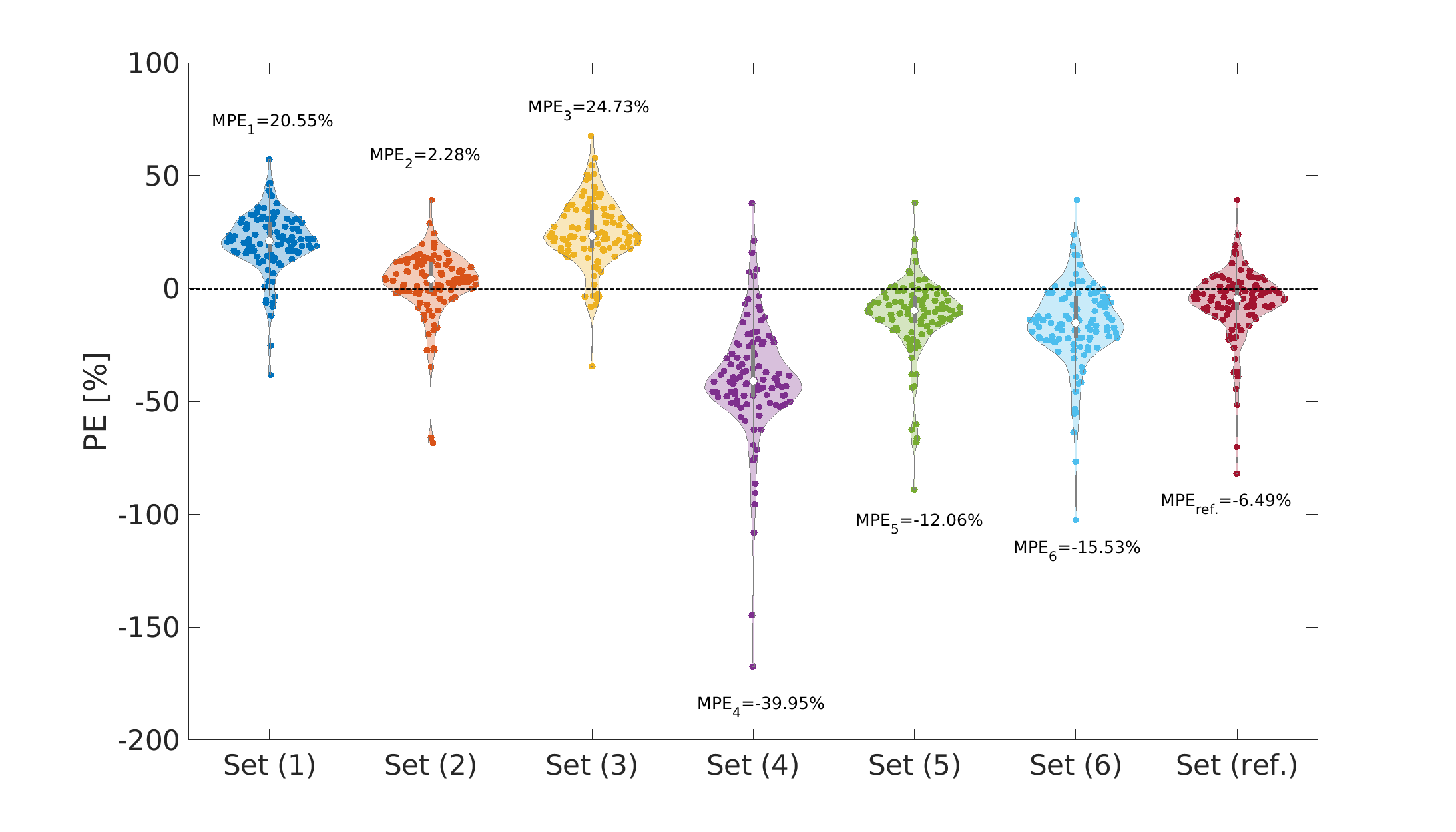}
\caption{
The violin plot visualization of the percentage error distributions of the measured and simulated atomic oxygen densities obtained in the simulations using the reactions ``Set (ref.)", ``Set (1)", ``Set (2)", ``Set (3)", ``Set (4)", ``Set (5)" and ``Set (6)" in figures \ref{fig:statistic_3_TALIF_noRie20}(b), \ref{fig:statistic_4_sou}(b), \ref{fig:statistic_4_sou}(d), \ref{fig:statistic_4_sou}(f), \ref{fig:statistic_5_sin}(b), \ref{fig:statistic_5_sin}(d) and \ref{fig:statistic_5_sin}(f), respectively. The violin plot is used to summarize the distributions of the percentage error provided in the aforementioned figures. For further details of the violin plot, see texts. 
}\label{fig:statistic_6_violin}
\end{figure}

The above-mentioned percentage error distributions of the measured and simulated atomic oxygen densities are summarized in figure \ref{fig:statistic_6_violin}, i.e. the distributions in the simulation results using the reactions ``Set (ref.)", ``Set (1)", ``Set (2)", ``Set (3)", ``Set (4)", ``Set (5)" and ``Set (6)" in figures \ref{fig:statistic_3_TALIF_noRie20}(b), \ref{fig:statistic_4_sou}(b), \ref{fig:statistic_4_sou}(d), \ref{fig:statistic_4_sou}(f), \ref{fig:statistic_5_sin}(b), \ref{fig:statistic_5_sin}(d) and \ref{fig:statistic_5_sin}(f), respectively. 
The summary is visualized with the violin plot, which combines the advantages of the above-mentioned box plot (see figure \ref{fig:statistic_2_TALIF}(b)) and histogram plot (see figure \ref{fig:statistic_3_TALIF_noRie20}(b)). Specifically, each violin consists of a white spot denoting the median, a gray bar representing the range of the lower and upper quartiles, and a gray line showing the region between the minimum and maximum. The width of the violin provides the relative distribution of the percentage error. It is intuitively and quantitatively observed that the percentage errors are shifted in the positive direction in the simulations using the reactions ``Set (1)", ``Set (2)" and ``Set (3)", while the percentage errors are shifted in the negative direction in the simulations using the reactions ``Set (4)", ``Set (5)" and ``Set (6)" compared to those using the reaction ``Set (ref.)". In other words, the simulation results in the case of removing a dominant atomic oxygen gain reaction channel in the simulations overall underestimate the measurement data, while the simulation results in the case of removing a dominant atomic oxygen loss reaction channel in the simulations overall overestimate the measurement data.

\section{Conclusion}
\label{sec:con}

The simulated atomic oxygen densities using the plug-flow model of this work are validated against the measured ones of the $\mu$APPJs reported in a range of publications \cite{Was10,Bib11,Wes16,Rie20, Mye21, Ste21, Ste22, Win22} using different measurement methods. These publications are from several research groups. These $\mu$APPJs are the COST-Jet \cite{Rie20, Mye21, Ste21, Ste22}, the COST-Jet prototype \cite{Was10,Bib11,Wes16}, and the $\mu$APPJ similar to the COST-Jet \cite{Win22}. It is worth to note that non-reproducibility of the measurement data due to the gas contamination and the absorbed power uncertainty was minimized by the COST-Jet \cite{Gol16}. The measurement data of the aforementioned $\mu$APPJs are as a function of the absorbed power (0.06 - 6.50 W), the He gas flow rate (200 - 1500 sccm), and the O$_2$ mixture ratio (0.1 - 2.0 $\%$). They were collected along the gas flow direction in the plasma channel region, at the middle or exit of the plasma channel region, and in the near effluent region. The measurement methods include the TALIF, the SEA and the OES. For further details of the above-mentioned information, see section \ref{sec:setup}.

The plug-flow model of this work is coupled with a Boltzmann equation under the two-term approximation to properly address the electron kinetics. 
It is observed in this work and in the literature \cite{Wil19,Mye21} that the atomic oxygen density at the middle of the plasma channel and in the near effluent are similar to those at the exit of the plasma channel. Therefore, only the plug-flow model calculation results at the exit of the plasma channel are used to compared with the measurement data which were collected at the middle or exit of the plasma channel region \cite{Mye21, Ste22, Win22}, and collected in the near effluent region \cite{Wes16,Rie20}, see section \ref{sec:mod}.
The plug-flow model calculation results along the gas flow direction in the plasma channel region are used to compared with the corresponding one dimensional spatially resolved measurement data \cite{Ste21,Win22}.
An agreement between the simulated atomic oxygen densities and the measured ones from most of the above-mentioned publications is obtained. A relatively good agreement is achieved between the simulations and most of the TALIF measurements. The simulated atomic oxygen densities overall slightly underestimate the SEA measurement data.

Our model prediction accuracy relative to the measurements from various research groups is quantified by the percentage error between the measured and simulated atomic oxygen densities. The SEA measurement data are overall slightly larger than the TALIF ones and our simulation results. These SEA measurement data are excluded in the investigation of the model prediction accuracy. A few TALIF measurement data are also excluded, since they are detected as significant outliers by the box plot of the percentage error. The percentage error between the remaining TALIF measurement data \cite{Wes16, Mye21, Ste21, Ste22} and our simulation results is visualized with the histogram plot to present its distribution intuitively. As expected, an approximate normal distribution is observed, and the mean percentage error is close to zero. This slight positive and negative deviation between the measurements and simulations is due to diverse factors such as model limitation, experimental error, and potential inconsistency between the input parameters of measurements and simulations.

The influence of removing one of the dominant atomic oxygen gain and loss reaction channels in the simulations on our model prediction accuracy relative to the above-mentioned TALIF measurements \cite{Wes16, Mye21, Ste21, Ste22} is analyzed. In the analysis, only the first three dominant gain and loss reaction channels at the plasma channel exit of the COST-Jet under a typical operating condition are considered for the sake of simplicity. The mean of the approximate normal distribution of the percentage error is shifted positively and negatively in the case of removing a dominant gain and loss reaction channel, which means the simulation results overall underestimate and overestimate the measurement data, respectively. These shifts are intuitively presented in a violin plot. The shift degree of the mean percentage error is not directly correlated with the contribution of the corresponding dominant reaction channel due to diverse factors: such as the different contribution of the dominant reaction channel at different plasma channel position under different operating condition, the number of the corresponding data points involved, and the convoluted chemical kinetics affected by an absence of the dominant reaction channel. However, the observations in this work still indicate that proper incorporation of the dominant reaction channels in the simulations is of importance for a promised accuracy of the model prediction relative to the measurements from various research groups. 

Overall, a framework has been presented via which simulation results can be quantitatively compared to measurement data under a variety of operating conditions. The distribution of the percentage error and the corresponding mean percentage error are intuitive metrics for comparing experiment and simulation, when there are a sufficient number of measurement data available. The COST-Jet, and other sources related to it, provide an ideal platform for such comparisons, due to their reproducibility and the wide range of measurement data that have been carried out on them by multiple research groups. Future work in this area may include the use of such comparisons for the optimization of chemical kinetics schemes based on minimization of the mean percentage error.

\section*{Acknowledgements}
Funded by the Deutsche Forschungsgemeinschaft (DFG, German Research Foundation) - Project-ID 327886311 (SFB 1316: A9).

\section*{ORCID iDs}
Youfan He  \href{target}{https://orcid.org/0000-0003-1275-7695}

Ralf Peter Brinkmann  \href{target}{https://orcid.org/0000-0002-2581-9894}

Efe Kemaneci  \href{target}{https://orcid.org/0000-0002-5540-0947}

Andrew R Gibson  \href{target}{https://orcid.org/0000-0002-1082-4359}

\clearpage

{\bf \LARGE Appendix}
\appendix

\section{Chemical kinetics considered in this work including the updates and supplements compared to those in previous work \cite{He21}}
\label{sec:chemkin}

\noindent 
\begin{ThreePartTable}
{\scriptsize
\setlength{\tabcolsep}{8pt}
\begin{longtable}[h]{p{0.7cm}p{6.5cm}p{5.5cm}p{1.5cm}p{0.2cm}}
\caption{
The helium only related volume reactions included in the $ \mathrm{He/O_2} $ model. For updates compared to our previous work \cite{He21} see notes below the table. The rate coefficient units are given in s$^{-1}$, m$^3$\:s$^{-1}$ and m$^6$\:s$^{-1}$ for one-, two- and three-body reactions, respectively. $\rm T_e$ is in eV and $\rm T_g$ in K, if not stated otherwise. The rate coefficient $ f(\epsilon) $ is taken from a look-up-table calculated via the referenced cross-section self-consistently coupled to the EEDF \cite{Tej19}. The reverse reaction rate coefficient of the electron-impact excitation labeled with a symbol ``*" near the number is calculated via the principle of {\em detailed balancing} \cite{LieBook2005}. 
}
\label{tab:ReactionListHe} \\

\hline
\# &Reaction     & Rate coefficient & Ref & \\ \hline
\\[\dimexpr-\normalbaselineskip+3pt]
\endfirsthead

\multicolumn{4}{c}
{{ \tablename\ \thetable{} -- $ Continued $ $ from $ $ previous $ $ page $}} \\
\hline
\# &Reaction     & Rate coefficient & Ref & \\ 
\hline
\\[\dimexpr-\normalbaselineskip+3pt]

\endhead

\hline \multicolumn{4}{r@{}}{{$ Continued $ $ on $ $ next $ $ page $}} \\
\\[\dimexpr-\normalbaselineskip+3pt]
\endfoot

\\[\dimexpr-\normalbaselineskip+3pt]
\endlastfoot

1   &   $ e + \mathrm{He} \rightarrow 2 e + \mathrm{He^+}  $   &   $ f(\epsilon) $   &     \cite{IST20230227,Alv14}     &      \\
\\[\dimexpr-\normalbaselineskip+3pt]
2$^*$   &   $ e + \mathrm{He} \rightarrow e + \mathrm{He(2 ^3S)}  $   &   $ f(\epsilon) $   &     \cite{IST20230227,Alv14}     &      \\
\\[\dimexpr-\normalbaselineskip+3pt]
3$^{"}$   &   $ e + \mathrm{He(2 ^3S)} \rightarrow 2 e + \mathrm{He^+}  $   &   $ f(\sigma) $   &     \cite{TRINITI-LXCat}$^{a}$     &      \\
\\[\dimexpr-\normalbaselineskip+3pt]
4$^{!}$   &   $ e + \mathrm{He_2^*} \rightarrow e + 2 \mathrm{He}  $   &   $ 3.8 \times 10^{-15} $   &     \cite{Sch18_PCCP,Del76}     &      \\
\\[\dimexpr-\normalbaselineskip+3pt]
5$^{"}$   &   $ e + \mathrm{He_2^*} \rightarrow 2 e + \mathrm{He_2^+}  $   &   $ f(\sigma) $   &     \cite{Fla81}$^{a}$     &      \\
\\[\dimexpr-\normalbaselineskip+3pt]

6$^{!"}$   &   $ e + \mathrm{He^+} \rightarrow \mathrm{He(2 ^3S)}  $   &   $ 6.76 \times 10^{-19} \: T{\rm _e}^{-0.5} $   &     \cite{Sta04,Bek66}     &      \\
\\[\dimexpr-\normalbaselineskip+3pt]
7$^{!"}$   &   $ 2 e + \mathrm{He^+} \rightarrow e + \mathrm{He(2 ^3S)}  $   &   $ 5.12 \times 10^{-39} \: T{\rm _e}^{-4.5} $   &     \cite{Sta04,Bek66}     &      \\
\\[\dimexpr-\normalbaselineskip+3pt]
8   &   $ e + \mathrm{He^+} + \mathrm{He} \rightarrow \mathrm{He(2 ^3S)} + \mathrm{He}  $   &   $ 7.4 \times 10^{-47} \: (T_{\rm e}/T_{\rm g})^{-2} $   &     \cite{Lie15}     &      \\
\\[\dimexpr-\normalbaselineskip+3pt]
9$^{!"}$   &   $ e + \mathrm{He_2^+} \rightarrow \mathrm{He} + \mathrm{He}  $   &   $ 1.0 \times 10^{-14} $   &     \cite{Lie15,Sta06}     &      \\
\\[\dimexpr-\normalbaselineskip+3pt]
10$^{!"}$   &   $ e + \mathrm{He_2^+} \rightarrow \mathrm{He(2 ^3S)} + \mathrm{He}  $   &   $ 8.9 \times 10^{-15} \: (T{\rm _e[K]}/T_{\rm g})^{-1.5} $   &     \cite{ Sak06,Bro05,Gol02_2}     &      \\
\\[\dimexpr-\normalbaselineskip+3pt]

11$^{!}$   &   $ \mathrm{He^+} + 2 \mathrm{He} \rightarrow \mathrm{He_2^+} + \mathrm{He}  $   &   $ 1.1 \times 10^{-43} $   &     \cite{Sak06,Bro05,Gol02_2}     &      \\
\\[\dimexpr-\normalbaselineskip+3pt]

12$^{!}$   &   $ \mathrm{He(2 ^3S)} + 2 \mathrm{He} \rightarrow \mathrm{He_2^*} + \mathrm{He}  $   &   $ 2 \times 10^{-46} $   &     \cite{Sak06,Bro05,Gol02_2}     &      \\
\\[\dimexpr-\normalbaselineskip+3pt]
13   &   $ \mathrm{He(2 ^3S)} + 2 \mathrm{He} \rightarrow 3 \mathrm{He}  $   &   $ 2 \times 10^{-46} $   &     \cite{Lie15}     &      \\
\\[\dimexpr-\normalbaselineskip+3pt]
14$^{!"}$   &   $ \mathrm{He(2 ^3S)} + \mathrm{He(2 ^3S)} \rightarrow e + \mathrm{He_2^+}  $   &   $ 2.03 \times 10^{-15} \: (T_{\rm g}/300)^{0.5}  $   &     \cite{Lie15,Wan06,Alv92}     &      \\
\\[\dimexpr-\normalbaselineskip+3pt]
15$^{!}$   &   $ \mathrm{He(2 ^3S)} + \mathrm{He(2 ^3S)} \rightarrow e + \mathrm{He^+} + \mathrm{He}  $   &   $ 8.7 \times 10^{-16} \: (T_{\rm g}/300)^{0.5} $   &     \cite{Lie15,Wan06,Alv92}     &      \\
\\[\dimexpr-\normalbaselineskip+3pt]
16$^{!}$   &   $ \mathrm{He(2 ^3S)} + \mathrm{He_2^*} \rightarrow e + \mathrm{He_2^+} + \mathrm{He}  $   &   $ 2.0 \times 10^{-15} $   &     \cite{Lie15,Del76}     &      \\
\\[\dimexpr-\normalbaselineskip+3pt]
17$^{!}$   &   $ \mathrm{He(2 ^3S)} + \mathrm{He_2^*} \rightarrow e + \mathrm{He^+} + 2 \mathrm{He}  $   &   $ 5.0 \times 10^{-16} $   &     \cite{Lie15,Del76}     &      \\
\\[\dimexpr-\normalbaselineskip+3pt]
18$^{!}$   &   $ \mathrm{He_2^*} \rightarrow 2 \mathrm{He}  $   &   $ 1 \times 10^4 $   &     \cite{Bro05,Gol02_2}     &      \\ 
\\[\dimexpr-\normalbaselineskip+3pt]
19   &   $ \mathrm{He_2^*} + \mathrm{He} \rightarrow 3 \mathrm{He}  $   &   $ 1.5 \times 10^{-21} $   &     \cite{Lie15}     &      \\
\\[\dimexpr-\normalbaselineskip+3pt]
20$^{!"}$   &   $ \mathrm{He_2^*} + \mathrm{He_2^*} \rightarrow e + \mathrm{He_2^+} + 2 \mathrm{He}  $   &   $ 1.2 \times 10^{-15} $   &     \cite{Lie15,Del76}     &      \\
\\[\dimexpr-\normalbaselineskip+3pt]
21$^{!}$   &   $ \mathrm{He_2^*} + \mathrm{He_2^*} \rightarrow e + \mathrm{He^+} + 3 \mathrm{He}  $   &   $ 3.0 \times 10^{-16} $   &     \cite{Lie15,Del76}     &      \\
\\[\dimexpr-\normalbaselineskip+3pt]

\hline
\end{longtable}}

\scriptsize
\begin{tablenotes}
\item[]$ ^{!} $ Compared to our previous work \cite{He21}, the ``Ref" are updated. 
\item[]$ ^{"} $ Compared to our previous work \cite{He21}, the ``Rate coefficient" is updated. 
\item[]$ ^{!"} $ Compared to our previous work \cite{He21}, the ``Rate coefficient" and the ``Ref" are updated. 
\item[]$ ^{a} \:\:$ The rate coefficient for cases marked with $ f(\sigma) $ is directly evaluated according to the calculated EEDF and the corresponding electron-impact cross-section from "Ref".
\end{tablenotes}

\end{ThreePartTable}

\noindent 
\begin{ThreePartTable}
{\scriptsize
\setlength{\tabcolsep}{8pt}
\begin{longtable}[h]{p{0.7cm}p{6.5cm}p{5.0cm}p{1.3cm}p{0.2cm}}
\caption{
The oxygen related volume reactions in the $ \mathrm{He/O_2} $ model. For updates compared to our previous work \cite{He21} see notes below the table. ``O$_2$" in this table represents O$_2(v=0)$. M is the background gas helium. The rate coefficient units are given in m$^3$\:s$^{-1}$ and m$^6$\:s$^{-1}$ for two- and three-body reactions, respectively. $\rm T_e$ is in eV and $\rm T_g$ in K, if not stated otherwise. The rate coefficient $ f(\epsilon) $ is taken from a look-up-table calculated via the referenced cross-section self-consistently coupled to the EEDF \cite{Tej19}. The reverse reaction rate coefficients of the electron-impact excitation labeled with a symbol ``*" near the number are calculated via the principle of {\em detailed balancing} \cite{LieBook2005}.
}
\label{tab:ReactionListHeO2} \\

\hline
\# &Reaction     & Rate coefficient & Ref & \\ \hline
\\[\dimexpr-\normalbaselineskip+3pt]
\endfirsthead

\multicolumn{4}{c}
{{ \tablename\ \thetable{} -- $ Continued $ $ from $ $ previous $ $ page $}} \\
\hline
\# &Reaction     & Rate coefficient & Ref & \\ 
\hline
\\[\dimexpr-\normalbaselineskip+3pt]

\endhead

\hline \multicolumn{4}{r@{}}{{$ Continued $ $ on $ $ next $ $ page $}} \\
\\[\dimexpr-\normalbaselineskip+3pt]
\endfoot

\\[\dimexpr-\normalbaselineskip+3pt]
\endlastfoot

\\[\dimexpr-\normalbaselineskip+3pt]

1   &   $ e + \mathrm{O(^3 P)} + \mathrm{O_2} \rightarrow \mathrm{O^-} + \mathrm{O_2}  $   &   $ 1 \times 10^{-43} $   &     \cite{CapBook}     &      \\
\\[\dimexpr-\normalbaselineskip+3pt]
2   &   $ e + \mathrm{O(^3 P)} + \mathrm{O_2} \rightarrow \mathrm{O_2^-} + \mathrm{O(^3 P)}  $   &   $ 1 \times 10^{-43} $   &     \cite{CapBook}     &      \\
\\[\dimexpr-\normalbaselineskip+3pt]
3$^* \: ^{"}$   &   $ e + \mathrm{O(^3 P)} \rightarrow e + \mathrm{O(^1 D)}  $   &   $ f(\sigma) $   &     \cite{IST20231031,Alv14}$^a$     &      \\
\\[\dimexpr-\normalbaselineskip+3pt]
4$ ^{"} $   &   $ e + \mathrm{O(^3 P)} \rightarrow 2 e + \mathrm{O^+}  $   &   $ f(\sigma) $   &     \cite{IST20231031,Alv14}$^a$     &      \\
\hline
\end{longtable}}

\scriptsize
\begin{tablenotes}
\item[]$ ^{!} \:\:$ Compared to our previous work \cite{He21}, the ``Ref" is updated. 
\item[]$ ^{"} $ Compared to our previous work \cite{He21}, the ``Rate coefficient" is updated. 
\item[]$ ^{!"} $ Compared to our previous work \cite{He21}, the ``Rate coefficient" and the ``Ref" are updated. 

\item[]$ ^{!^D} $ Compared to our previous work \cite{He21}, the ``Rate coefficient" and the ``Ref" are updated based on the work of Dias $et$ $al$ \cite{Dia23}. For Reactions 47, 115, 118, 119 and 124, only the ``Ref" is updated.

\item[]$ ^{a} \:\:$ The rate coefficient for cases marked with $ f(\sigma) $ is directly evaluated according to the calculated EEDF and the corresponding electron-impact cross-section from "Ref".

\item[]$ ^{b} \:\:$ Reaction 125: $ \mathrm{O_2 (a^1 \Delta_g)} + \mathrm{O_3} \rightarrow \mathrm{O(^1 D)} + 2 \mathrm{O_2}  $ with a rate coefficient of $ 1.01 \times 10^{-17} $ m$^3$\:s$^{-1}$ that was used in \cite{He21} (P. 24) is deleted in the current work. 
The rate coefficient of Reaction 125 is much larger than that of a similar reaction channel Reaction 124, e.g. the former (producing $\mathrm{O(^1 D)}$) is around 649 times larger than the latter (producing $\mathrm{O(^3 P)}$) for a gas temperature of 350 K. Moreover, Reaction 124 with the same rate coefficient is included in the studies containing oxygen chemistry \cite{Sta04,Tur15,Gue19_rev,Liu21_2,Bri21,Dia23}, while Reaction 125 is not considered in these studies. It should be emphasized that well agreement between the measured and simulated $\mathrm{O_2 (a^1 \Delta_g)}$ densities is achieved in the work of Dias $et$ $al$ \cite{Dia23}, and worse agreement can be obtained in the case of considering Reaction 125 with the aforementioned rate coefficient in their simulations due to the significant $\mathrm{O_2 (a^1 \Delta_g)}$ loss. Therefore, Reaction 125 is excluded from the current work.
\end{tablenotes}

\end{ThreePartTable}

\begin{ThreePartTable}
{\scriptsize
\setlength{\tabcolsep}{8pt}
\begin{longtable}[h]
{p{0.2cm}p{6.2cm}p{5.4cm}p{2.2cm}p{0.1cm}}
\caption{
The additionally considered volume reactions in the $\mathrm{He/O_2}$ model compared to our previous work \cite{He21}. Reaction 1-4 are additionally considered due to the inclusion of $\mathrm{O_2 (b^1 \Sigma_{g^+})}$ in the current chemical model. Reaction 5-16 and 17-87 are additionally considered in this study based on the work of Dias $et$ $al$ \cite{Dia23} and Brisset $et$ $al$ \cite{Bri21}, respectively. ``O$_2$" in this table represents O$_2(v=0)$. The rate coefficient units are given in m$^3$\:s$^{-1}$ and m$^6$\:s$^{-1}$ for two- and three-body reactions, respectively. $\rm T_e$ is in eV, and $\rm T_g$ is in K. The rate coefficient for cases marked with $ f(\epsilon) $ is taken from a look-up-table calculated via the referenced cross-section self-consistently coupled to the EEDF \cite{Tej19}. The reverse reaction rate coefficients of the electron-impact excitation labeled with a symbol ``*" near the number are calculated via the principle of {\em detailed balancing} \cite{LieBook2005}. For the ``Reaction" labeled with a symbol ``$ \sim $" near the number, the ``Rate coefficient" of the electronically excited state is assumed to be the same as that of the corresponding lower state or that of O$_2(v=0)$. 
}
\label{tab:ReactionListHeO2updated} \\

\hline
\# &Reaction     & Rate coefficient & Ref & \\ \hline
\\[\dimexpr-\normalbaselineskip+3pt]
\endfirsthead

\multicolumn{4}{c}
{{ \tablename\ \thetable{} -- $ Continued $ $ from $ $ previous $ $ page $}} \\
\hline
\# &Reaction     & Rate coefficient & Ref & \\ 
\hline
\\[\dimexpr-\normalbaselineskip+3pt]

\endhead

\hline \multicolumn{4}{r@{}}{{$ Continued $ $ on $ $ next $ $ page $}} \\
\\[\dimexpr-\normalbaselineskip+3pt]
\endfoot

\\[\dimexpr-\normalbaselineskip+3pt]
\endlastfoot

&  \textbf{\underline{R1 - R4 from IST-Lisbon database}}  \cite{IST20231031,Alv14}   &      &          &      \\
\\[\dimexpr-\normalbaselineskip+3pt]

1$^*$   &   $ e + \mathrm{O_2} \rightarrow e + \mathrm{O_2 (b^1 \Sigma_{g^+})}  $   &   $ f(\epsilon) $   &     \cite{IST20231031,Alv14}     &      \\
\\[\dimexpr-\normalbaselineskip+3pt]
2$^*$   &   $ e + \mathrm{O_2 (a^1 \Delta_g)} \rightarrow e + \mathrm{O_2 (b^1 \Sigma_{g^+})}  $   &   $ f(\sigma) $   &     \cite{IST20231031,Alv14}$^a$     &      \\
\\[\dimexpr-\normalbaselineskip+3pt]
3   &   $ e + \mathrm{O_2 (b^1 \Sigma_{g^+})} \rightarrow e + \mathrm{O(^3 P)} + \mathrm{O(^3 P)}  $   &   $ f(\sigma) $   &     \cite{IST20231031,Alv14}$^a$     &      \\
\\
\hline
\end{longtable}}

\scriptsize
\begin{tablenotes}
\item[]$ ^{a} \:\:$ The rate coefficient for cases marked with $ f(\sigma) $ is directly evaluated according to the calculated EEDF and the corresponding electron-impact cross-section from "Ref".
\end{tablenotes}

\end{ThreePartTable}

\begin{ThreePartTable}
{\scriptsize
\setlength{\tabcolsep}{8pt}
\begin{longtable}[h]{p{0.2cm}p{3.5cm}p{3.5cm}p{3.5cm}p{1.9cm}p{0.1cm}}
\caption{
The additionally considered electron-impact excitation reactions in the $\mathrm{He/O_2}$ model compared to our previous work \cite{He21}. 
These reactions represent the excitation from ground state to vibrationally excited states and high electronically excited states, which belongs to the helium \cite{IST20230227,Alv14} or oxygen \cite{IST20231031,Alv14} complete set of the IST-Lisbon database yielding good agreement between calculated and measured swarm parameters. Note that these vibrationally excited states and high electronically excited states are not considered as species in our $\mathrm{He/O_2}$ model. In other words, the production and destruction of these states (see equation (1) in \cite{He21}) are not considered. Only the energy loss due to these excitation reactions (see equation (11) in \cite{He21}) is taken into account based on the corresponding threshold values given in table \ref{tab:ReactionListHeO2updated_loki}. For the sake of clarity, only the threshold values in the complete set of the IST-Lisbon database \cite{IST20230227,IST20231031,Alv14} are presented while the corresponding state notations are omitted.
``O$_2$" in this table represents O$_2(v=0)$. The rate coefficient for cases marked with $ f(\epsilon) $ is taken from a look-up-table calculated via the referenced cross-section self-consistently coupled to the EEDF \cite{Tej19}.  
}
\label{tab:ReactionListHeO2updated_loki} \\

\hline
\# &Excitation  & Energy threshold (eV)   & Rate coefficient & Ref & \\ \hline
\\[\dimexpr-\normalbaselineskip+3pt]
\endfirsthead

\multicolumn{5}{c}
{{ \tablename\ \thetable{} -- $ Continued $ $ from $ $ previous $ $ page $}} \\
\hline
\# &Excitation  & Energy threshold (eV)   & Rate coefficient & Ref & \\ 
\hline
\\[\dimexpr-\normalbaselineskip+3pt]

\endhead

\hline \multicolumn{5}{r@{}}{{$ Continued $ $ on $ $ next $ $ page $}} \\
\\[\dimexpr-\normalbaselineskip+3pt]
\endfoot

\\[\dimexpr-\normalbaselineskip+3pt]
\endlastfoot

1   &   $ e + \mathrm{He} \rightarrow e + \mathrm{He}  $   &   20.62   &   $ f(\epsilon) $   &     \cite{IST20230227,Alv14}     &      \\
\\[\dimexpr-\normalbaselineskip+3pt]
2   &   $ e + \mathrm{He} \rightarrow e + \mathrm{He}  $   &   20.96   &   $ f(\epsilon) $   &     \cite{IST20230227,Alv14}     &      \\
\\[\dimexpr-\normalbaselineskip+3pt]
3   &   $ e + \mathrm{He} \rightarrow e + \mathrm{He}  $   &   21.218   &   $ f(\epsilon) $   &     \cite{IST20230227,Alv14}     &      \\
\\[\dimexpr-\normalbaselineskip+3pt]
4   &   $ e + \mathrm{He} \rightarrow e + \mathrm{He}  $   &   22.719   &   $ f(\epsilon) $   &     \cite{IST20230227,Alv14}     &      \\
\\
\hline
\end{longtable}}

\scriptsize
\begin{tablenotes}
\item[]$ ^{a} \:\:$ The rate coefficient for cases marked with $ f(\sigma) $ is directly evaluated according to the calculated EEDF and the corresponding electron-impact cross-section from "Ref".
\end{tablenotes}

\end{ThreePartTable}

\begin{table}[hbt!]\scriptsize
\setlength{\tabcolsep}{25pt}
\centering
\caption{The electron-impact elastic collisions in the $\mathrm{He/O_2}$ model. For updates compared to our previous work \cite{He21} see notes below the table. ``O$_2$" in this table represents O$_2(v=0)$. 
}
\begin{tabular}{lllll}
\\[\dimexpr-\normalbaselineskip+3pt]
\hline
\\[\dimexpr-\normalbaselineskip+3pt]
\# &Collision     &  Ref  & \\ \hline
\\[\dimexpr-\normalbaselineskip+3pt]
1  & $ e + \mathrm{He} \rightarrow e + \mathrm{He} $ &   \cite{IST20230227,Alv14} &      \\ 
\\[\dimexpr-\normalbaselineskip+3pt]
4  & $  e+ \mathrm{O(^3 P)}  \rightarrow e + \mathrm{O(^3 P)}  $ &   \cite{IST20231031,Alv14}   &     \\ 
\\[\dimexpr-\normalbaselineskip+3pt]
5$^{!}$  & $  e+ \mathrm{O_2}  \rightarrow e + \mathrm{O_2}  $ &   \cite{IST20231031,Alv14} &      \\ 
\\[\dimexpr-\normalbaselineskip+3pt]

\hline 
\end{tabular}

\begin{tablenotes}
\tiny
\item[1]$ ^{!} \:\:$ Compared to our previous work \cite{He21}, the ``Ref" is updated. Specifically, the cross-section data of reaction 5 in ``Ref" \cite{IST20231031,Alv14} are used in the current work instead of those in ``Ref" \cite{He21}.
\item[2]$ ^{a} \:\:$ The “Reaction” containing nitrogen species in \cite{He21} (i.e. reactions 2-3 and 7-9) are deleted. 
\item[3]$ ^{b} \:\:$ The “Reaction” (i.e. reaction 6) is deleted due to an improved treatment of the elastic collision cross-section data. 
\end{tablenotes}

\label{tab:elaHeO2}
\end{table}

\begin{table}[H]\scriptsize
\setlength{\tabcolsep}{14pt}
\centering
\caption{The neutral wall reactions in the $\mathrm{He/O_2}$ model. For updates compared to our previous work \cite{He21} see notes below the table. ``O$_2$" in this table represents O$_2(v=0)$. }
\begin{tabular}{lllll}
\\[\dimexpr-\normalbaselineskip+3pt]
\hline
\\[\dimexpr-\normalbaselineskip+3pt]
\# &Reaction     &  Probability($\gamma$) &  Ref &  \\ \hline
\\[\dimexpr-\normalbaselineskip+3pt]
1  & $ \mathrm{He(2 ^3S)} + \mathrm{wall} \rightarrow \mathrm{He}  $ & $1$ &    \cite{Liu10_FromEfe,Yan16} &    \\ 
\\[\dimexpr-\normalbaselineskip+3pt]
2  & $ \mathrm{He_2^*} + \mathrm{wall} \rightarrow 2 \mathrm{He}  $ & $1$ &    \cite{Liu10_FromEfe,Yan16} &    \\ 
\\[\dimexpr-\normalbaselineskip+3pt]
7  & $ \mathrm{O(^1D)} + \mathrm{wall} \rightarrow \mathrm{O(^3P)}  $ & $1$  &    \cite{Gor95}$ ^d $   & \\ 
\\[\dimexpr-\normalbaselineskip+3pt]
8  & $ \mathrm{O_2(a^1 \Delta_g)} + \mathrm{wall} \rightarrow \mathrm{O}_2  $ & $0.00022$ &   \cite{Boo20}$ ^d $  &   \\ 
\\[\dimexpr-\normalbaselineskip+3pt]
14  & $ \mathrm{O(^3P)} + \mathrm{wall} \rightarrow 1/2\mathrm{O}_2  $ & $0.002$  &  estimated from \cite{Boo20}$ ^e $   &  \\ 
\\[\dimexpr-\normalbaselineskip+3pt]
15  & $ \mathrm{O_2 (b^1 \Sigma_{g^+})} + \mathrm{wall} \rightarrow \mathrm{O}_2  $ & $0.135$ &   \cite{Boo22}$ ^e $  &   \\ 
\\[\dimexpr-\normalbaselineskip+3pt]

\\[\dimexpr-\normalbaselineskip+3pt]
\hline
\end{tabular}

\begin{tablenotes}
\tiny
\item[1]$^a$ The “Reaction” containing nitrogen species in \cite{He21} (i.e. reactions 3-6 and 11) are deleted. 
\item[2]$ ^b $ The “Reaction” which is included for a test of the sensitivity analysis in \cite{He21} (i.e. reactions 9, 10 and 13) are deleted.
\item[3]$ ^c $ The “Reaction” containing O$_2(v>0)$ (i.e. reaction 12) is deleted due to the exclusion of vibrationally excited molecules from the current work.
\item[4]$ ^d $ Compared to our previous work \cite{He21}, the probabilities of reactions 7 and 8 are updated in this study to be consistent with those of Dias $et$ $al$ \cite{Dia23}.
\item[5]$ ^e $ Reactions 14 and 15 are additionally considered in this study based on the work of Dias $et$ $al$ \cite{Dia23}. 
The probabilities of reaction 14 as a function of pressure, discharge current and wall temperature reported in \cite{Dia23} are estimated to a constant value in this study. This constant is around the mode value of the probabilities reported in \cite{Dia23}.
\end{tablenotes}

\label{tab:wrHeO2}
\end{table}

\newpage
\section{Comparison between the simulation results in this work and the measurement data of the He/O$_2$ $\mu$APPJs considered in previous work \cite{He21}}
\label{sec:He21APPJ}

\begin{figure}[h!]
\begin{minipage}[b]{0.5\textwidth}
\includegraphics[width=0.95\textwidth]{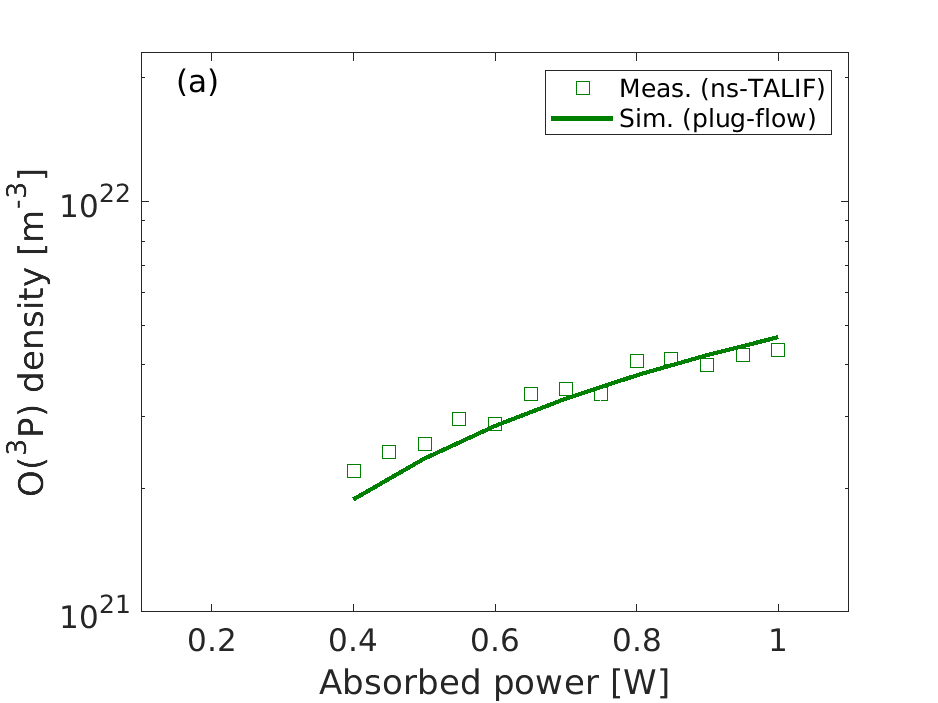} \\ 
\end{minipage}
\begin{minipage}[b]{0.5\textwidth}
\includegraphics[width=0.95\textwidth]{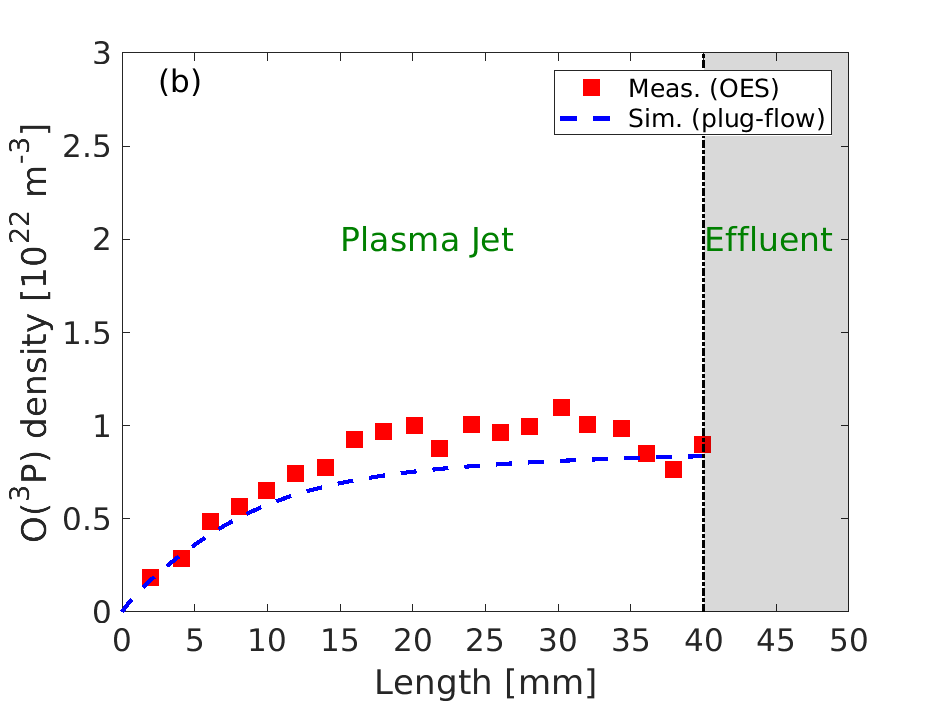} \\ 
\end{minipage}
\caption{
(a) The atomic oxygen density at the middle of the plasma channel measured with the ns-TALIF method in a He/$\mathrm{O_2}$ $\rm \mu$APPJ by Waskoenig $et$ $al$ \cite{Was10} (\textcolor{OliveGreen}{$\square$}) together with the plug-flow model calculation results at the middle of the plasma channel in this work (\textcolor{OliveGreen}{\full}) for a variation of the absorbed power from 0.40 W to 1.00 W.  
(b) The spatially resolved atomic oxygen density measured with the OES method in the gas flow direction of a He/$\mathrm{O_2}$ $\rm \mu$APPJ by Bibinov $et$ $al$ \cite{Bib11} (\textcolor{red}{$\blacksquare$}) and the corresponding plug-flow model calculation results in this work (\textcolor{blue}{\dashed}). More details regarding the $\mu$APPJs such as the operating conditions and the measurement methods, see section \ref{sec:setup}.
}\label{fig:Was10Bib11}
\end{figure}

The measurement data of the He/O$_2$ $\mu$APPJs by Waskoenig $et$ $al$ \cite{Was10} and Bibinov $et$ $al$ \cite{Bib11} considered in previous work \cite{He21} and the corresponding simulated atomic oxygen densities in this work using the plug-flow model in section \ref{sec:mod} and the chemical kinetics in section \ref{sec:Chem_kin} are presented in figure \ref{fig:Was10Bib11}. More details regarding the $\mu$APPJs such as the operating conditions and the measurement methods, see section \ref{sec:setup}. A good agreement between the measurement data and the simulation results is similarly obtained in this work relative to those comparisons in previous work \cite{He21}. However, due to the assumption of the $5\%$ power transfer efficiency in the simulations of the aforementioned $\mu$APPJs (see section \ref{sec:setup}), these calculation results in figure \ref{fig:Was10Bib11} are only used as a comparison with those in previous study \cite{He21}, and only considered in \ref{sec:He21APPJ}. Note that these results are not analyzed in section \ref{sec:res}.

\newpage
\clearpage
\bibliographystyle{unsrt}
\bibliography{ref}

\end{document}